\documentclass[12pt]{amsart}
\usepackage{amsbsy,amssymb,amscd,amsfonts,latexsym,amstext,delarray,
amsmath,graphicx, xcolor, mathtools, float, tikz, bbm} 
\usetikzlibrary{decorations.pathreplacing,calligraphy, patterns, calc,fit,matrix,arrows,automata,positioning, patterns,decorations.markings, decorations.pathmorphing}
\usepackage[font={small, sc}]{caption}
\input xypic

\usepackage{color}

\newtheorem{thm}{Theorem}[section]
\newtheorem{prop}[thm]{Proposition}
\newtheorem{cor}[thm]{Corollary}
\newtheorem{lem}[thm]{Lemma}

\newtheorem{defn}[thm]{Definition}
\newtheorem{rem}[thm]{Remark}

\numberwithin{equation}{section}

\def\bH{{\mathbb H}}

\def\bP{{\mathbb P}}
\def\bQ{{\mathbb Q}}
\def\bR{{\mathbb R}}

\def\bZ{{\mathbb Z}}

\def\A{{\mathbb A}}
\def\C{{\mathbb C}}

\renewcommand{\H}{{\mathbb H}}
\def\N{{\mathbb N}}
\renewcommand{\P}{{\mathbb P}}
\def\Q{{\mathbb Q}}
\def\R{{\mathbb R}}
\def\Z{{\mathbb Z}}
\def\K{{\mathbb K}}

\def\cA{{\mathcal A}}
\def\cB{{\mathcal B}}

\def\cD{{\mathcal D}}
\def\cE{{\mathcal E}}

\def\cG{{\mathcal G}}
\def\cH{{\mathcal H}}
\def\cI{{\mathcal I}}

\def\cL{{\mathcal L}}
\def\cM{{\mathcal M}}

\def\cP{{\mathcal P}}

\def\cS{{\mathcal S}}

\def\cU{{\mathcal U}}

\def\cW{{\mathcal W}}

\def\GL{{\rm GL}}

\def\PGL{{\rm PGL}}
\def\PSL{{\rm PSL}}

\def\SL{{\rm SL}}

\def\Sp{{\rm Spec}}

\def\Tr{{\rm Tr}}

\title[QSM and the boundary of modular curves]{Quantum Statistical Mechanics and the Boundary of Modular Curves}
\author{Matilde Marcolli and Jane Panangaden}
\date{2023}
\address{Department of Mathematics and Department of Computing
and Mathematical Sciences, California Institute of Technology, Pasadena \\ USA}
\email{matilde@caltech.edu}
\address{Mathematics Department, Pitzer College, Claremont \\ USA}
\email{jane\_panangaden@pitzer.edu}

\begin{document}
\maketitle

\begin{abstract}
The theory of limiting modular symbols provides a noncommutative geometric
model of the boundary of modular curves that includes irrational points in addition
to cusps. A noncommutative space associated to this boundary is constructed, as
part of a family of noncommutative spaces associated to different continued fractions
algorithms, endowed with the structure of a quantum statistical mechanical system. 
Two special cases of this family of quantum systems can
be interpreted as a boundary of the system associated to the
Shimura variety of $\GL_2$ and an analog for $\SL_2$. The structure of KMS
states for this family of systems is discussed. In the geometric cases, the ground 
states evaluated on boundary arithmetic elements are given by pairings of cusp 
forms and limiting modular symbols. 
\end{abstract}

\section{Introduction}

Since the introduction of the Bost--Connes system in the mid '90s, \cite{BC}, a very rich interplay
between number theory and quantum statistical mechanics developed, involving the Galois theory
of abelian and non-abelian extensions of number fields 
(\cite{CoMaRa}, \cite{Corn1}, \cite{HaPa}, \cite{LLN}, \cite{ManMar3}, \cite{Yalk}),
Shimura varieties (\cite{CoMa}, \cite{HaPa}, \cite{LLN2}), $L$-series and zeta functions 
(\cite{CCM}, \cite{Corn2}), etc. On the other hand, the work of Manin and Marcolli, \cite{ManMar},
and the subsequent work \cite{KesStra}, \cite{ManMar2}, \cite{Mar}, developed a theory of
limiting modular symbols, and a related noncommutative geometry model of the boundary of
modular curves.  It was remarked in \S 7.9 of \cite{CoMa-book}, as well as in \cite{CMR2},
that the noncommutative
compactification of modular curves of \cite{ManMar} should fit as a ``boundary stratum" of
the quantum statistical mechanical system of \cite{CoMa} that generalized the Bost--Connes
system from the case of $\GL_1$ to the case of $\GL_2$ and has the geometry of modular curves
directly built into its construction through the Shimura variety of $\GL_2$. However, the relation 
between the limiting modular symbols of \cite{ManMar} and the 
$\GL_2$ quantum statistical mechanical system of \cite{CoMa} has not been fully 
analyzed. It is the purpose of this paper to describe a construction of a ``boundary algebra"
for the $\GL_2$ system, endowed with an induced time evolution, that is built on the
noncommutative boundary of modular curves described in \cite{ManMar} and on 
limiting modular symbols.

\smallskip

The $\GL_2$ system of \cite{CoMa} is based on a convolution algebra involving Hecke operators
and possibly degenerate level structures on modular curves. More precisely,  
in  \cite{CoMa} one considers functions on the set 
$$ \{ (g,\rho,z) \in \GL_2^+(\Q) \times M_2(\hat \Z) \times \H \,|\, g\rho \in M_2(\hat \Z) \} $$
that are {\em invariant} under the action of $\SL_2(\Z)\times \SL_2(\Z)$ by
$$ (\gamma_1,\gamma_2): (g,\rho,z) \mapsto (\gamma_1 g \gamma_2^{-1}, \gamma_2 \rho, \gamma_2 z). $$
This algebra is then endowed with a time evolution and covariant representations.

\smallskip 

The $\GL_2$ system has partition function given by $\zeta(\beta)\zeta(\beta-1)$ and a symmetry group 
$\textrm{GL}_2(\mathbb{A}_f)/\mathbb{Q}^*\simeq {\rm Aut}(F)$, where $F$ is the modular field. These
symmetries include both automorphisms and endomorphisms of the algebra of observables, compatible
with the Hamiltonian time evolution. The $\beta$-KMS states exhibit symmetry breaking at two critical temperatures, $\beta=1$ and $\beta=2$, with the extremal low-temperature states being parameterized by invertible $\bQ$-lattices up to scaling, which can be seen equivalently as the set 
$\GL_2(\Q)\backslash \GL_2(\A)/\C^*$ of points of the Shimura variety of $\GL_2$. This is
a direct generalization of the original Bost--Connes case, where similarly the extremal low-temperature
KMS states can be identified with the points $\GL_1(\Q)\backslash \GL_1(\A)/\R^*_+$ of the
Shimura variety of $\GL_1$ (see \cite{CMR2} for a broader discussion of this viewpoint). 
In the $\GL_2$ system, the ground states, when evaluated on points in the arithmetic algebra, yield generators of specializations of the modular field to points in the upper half plane. The Shimura varieties viewpoint was
further developed in \cite{Ab}, \cite{HaPa}, \cite{Shen}. 

\smallskip 

This general viewpoint includes extensions of the original Bost--Connes system from $\Q$ to 
arbitrary number fields $K$. Such systems were first introduced by Ha and Paugam 
in \cite{HaPa}, in the context of
the general setting for Shimura varieties, and by  Connes, Marcolli, and Ramachandran  
in \cite{CoMaRa}, \cite{CMR2} in the case of complex 
multiplication, where one obtains a quantum statistical mechanical interpretation of the explicit class field
theory for imaginary quadratic fields. In \cite{LLN} the KMS states analysis was extended
to the systems for arbitrary number fields, and in \cite{Yalk} a construction of an arithmetic
subalgebra was obtained for these systems. The Bost--Connes systems
for number fields were also used in \cite{Corn1}, \cite{Corn2} to obtain new number theoretic 
reconstruction results for number fields. 

\smallskip

The formulation of \cite{CoMaRa}, for imaginary quadratic fields, is based upon geometric objects given by $1$-dimensional $K$-lattices, which take on the role of the $\bQ$-lattices of
the Bost--Connes and the $\GL_2$ case. The resulting $C^*$-dynamical system is known as the CM system, due to its connection with complex multiplication. The extremal zero-temperature KMS states evaluated on arithmetic points are related to $\mathbb{A}^*_{K,f}/K^*$ for an imaginary quadratic field $K$. The CM system is thus connected to the explicit class field theory for imaginary quadratic extensions. The CM system can be viewed as a specialization of the $\GL_2$ system, since a $K$-lattice can be viewed as a 2-dimensional $\bQ$-lattice. The construction using $K$-lattices has been extended by Laca, Larsen, and Neshveyen in \cite{LLN} to all number fields. The construction yields some of the desired properties: the correct partition function, KMS-states, symmetries, and symmetry breaking behavior. The construction of an arithmetic algebra for the 
evaluation of the ground states and the action of Galois symmetries, 
which is a crucial part of the construction in relation to the quantum statistical mechanics approach to
the Hilbert's 12th problem, was obtained for arbitrary number fields by Yalkinoglu in \cite{Yalk}. 

\smallskip

In this work, we take a different approach. Instead of using the $K$-lattices, we view $\bP^1(\bR)$ as an ``invisible'' boundary of $\bH$, with points in $\bR$ representing pseudolattices which can be viewed as degenerations of complex tori as suggested by Manin's real multiplication program, \cite{Man}, \cite{Man2}. We construct a boundary version of the $\GL_2$-system by incorporating the boundary $\bP^1(\bR)$ directly, with an action of the shift operator which implements the shift on the continued fraction expansion. This approach is
motivated by the following observation. In \cite{ManMar}, \cite{ManMar2}, the ``noncommutative quotient"
of $\bP^1(\bR)$ by the action of $\PGL_2(\Z)$ by fractional linear transformations is seen as the 
noncommutative moduli space of noncommutative tori up to Morita equivalences, and the 
noncommutative tori are interpreted as non-algebro-geometric degenerations of elliptic curves as
the modulus $\tau \in \H$ approaches an irrational points of the boundary $\bP^1(\R)=\partial \H$.
In \cite{CMR2} and \cite{CoMa-book} it is observed that one can identify, 
as mentioned above, the set of 
low temperature extremal KMS states of the $\GL_2$ system with the set of classical points
$$ Sh(\GL_2,\H^\pm)=\GL_2(\Q)\backslash \GL_2(\A)/\C^* =\GL_2(\Q)\backslash (\GL_2(\A_f)\times \bH^\pm) $$ of the Shimura variety of $\GL_2$, by identifying this with the set 
$$ \SL_2(\Z) \backslash (\GL_2(\hat \Z) \times \H) $$
of invertible $2$-dimensional $\Q$-lattices (see \S I.17 of \cite{Mum}). 
The algebra of observables of the $\GL_2$ system (a noncommutative space) can then be seen as an additional
``non-classical" but rather ``quantum" part of this same Shimura variety. The noncommutativity in this
case arises from the action of $\GL_2(\Q)$ on $M_2(\A)$ rather than on $\GL_2(\A)$, hence
it represents degenerations of the level structure rather than of the elliptic curve itself. Namely,
the $\GL_2$-system can be seen (as in \cite{CMR2}) as the non-commutative space
$$ Sh^{nc}(\GL_2,\H^\pm)=\GL_2(\Q)\backslash (M_2(\A_f)  \times \H^\pm) $$
where the role of the extremal KMS states is to extract the classical points. The role of 
these classical and quantum parts was further discussed, for the original Bost--Connes
system in \cite{CCM2}. There is a natural way to simultaneously allow for both of these 
possible degenerations, of the elliptic curve (of a lattice to a pseudolattice) and of the
level structure (of an invertible labeling of the torsion points to a non-invertible one).
As suggested in \cite{CMR2} and \S 7.9 of \cite{CoMa-book}, 
this is achieved by considering the noncommutative space
$$ \overline{Sh^{nc}}(\GL_2,\H^\pm)=\GL_2(\Q)\backslash (M_2(\A_f)\times \bP^1(\C))\, , $$
which should be regarded as the natural noncommutative compactification of the
Shimura variety of $\GL_2$. The question then is how to appropriately interpret this noncommutative
space, by constructing an algebra of observables and a quantum statistical mechanical system that
will describe this boundary compactification of the $\GL_2$-system and that will also exhibit the
relation with the theory of limiting modular symbols and the properties of the non-commutative
boundary described in \cite{ManMar}, \cite{ManMar2}. This is the purpose of the present paper. 

\smallskip

A first difficulty in extending the $\GL_2$-system to the boundary is that on
$\bP^1(\R)=\partial \H$ the action of $\Gamma = \SL_2(\Z)$ by fractional linear transformation has dense orbits.
In the original construction of the $\GL_2$-system in \cite{CoMa} one considers the
space of $2$-dimensional $\Q$-lattices, which is described by the quotient 
$\Gamma \backslash (M_2(\hat\Z) \times \H)$, with the action by fractional linear transformations on $\H$,
and by a convolution algebra (which implements the commensurability relation) associated to the quotient of
$\GL_2^+(\Q)\times M_2(\hat\Z) \times \H$ by the action $(g,\rho,z) \mapsto (\gamma_1 g \gamma_2^{-1} \gamma_2 \rho, \gamma_2(z))$ of $\Gamma\times \Gamma$. 
When $\H$ is compactified with boundary $\bP^1(\R)=\partial \H$, these actions no longer
have a good classical quotient. This is indeed the source of the noncommutativity of this
boundary. This means that it no longer makes sense to consider functions that are invariant under this action. 
Thus, what we consider here, as an alternative, is to retain the $\SL_2(\Z)$ invariance as
above in the variables $(g,\rho)$, while replacing the action of $\SL_2(\Z)$ by fractional linear transformations
on $\bP^1(\R)$ (that extends the action on $z\in \H$) 
by a different action. Instead of requiring invariance, we introduce, as part of the algebra,
generators that implement this action (which amounts to taking a quotient in the noncommutative sense). 
The action considered is built out of the partial inverses of the shift of the continued fraction
expansion. (Regarding the resulting decoupling of this action and the
$\GL_2^+(\Q)$-action, see the comments below about isogenies.) 

\smallskip

We first extend the original definition of the $\GL_2$ system of \cite{CoMa} to other
subgroups of $\GL_2(\Q)$ that include the case of $\SL_2(\Z)$ and $\GL_2(\Z)$. The
main requirement for such subgroups $\Gamma \subset \GL_2(\Q)$ is to have an
associated Hecke algebra $\Xi = \Gamma \backslash \GL_2(\Q) /\Gamma$. In order to
account for some additional structure considered in the setting of limiting modular
symbols in \cite{ManMar}, we also introduce the choice of a finite index subgroup $G\subset \Gamma$
and the coset spaces $\bP_\alpha=\Gamma \alpha G/G$. 

\smallskip

We focus in particular on a choice of $\Gamma=\Gamma_N$, dependent on an integer 
$N \in \Z\smallsetminus \{ 0 \}$, consisting of matrices in $\GL_2(\Q)$ with determinant
in the subgroup of $\GL_1(\Q)$ generated by the prime factors of $N$ and $-1$.  
The main motivation for this choice is that 
these subgroups contain certain semigroups
associated to an $N$-dependent family of continued fraction algorithms that we will use in
the construction of a corresponding family of ``boundary systems".
This family includes a $\GL_2(\Z)$-version of the original Connes--Marcolli $\GL_2$-system as a special case. 
Moreover, the partition function for these systems has a natural interpretation as the
zeta function of the $\GL_2$ system (the zeta function of $\bP^1$) with a finite number
of Euler factors removed. 

\smallskip

We then construct two families of noncommutative algebras (both dependent on
the integer parameter $N$.) The first is a family of ``bulk algebras" that generalize 
the $\GL_2$ system of \cite{CoMa},
involving the Hecke algebra $\Xi_N$ with a (partially defined) action on the level structure 
$\rho\in M_2(\hat\Z)$, a discrete space $\tilde \cP_N$ built from the coset 
spaces $\bP_\alpha$, and the half planes $\H^\pm$.
The other is a corresponding family of ``boundary algebras" that are semigroup crossed products 
of the algebra of continuous functions from a ``disconnection" $\cD_{[0,1]\cap \Q}$ 
of the interval $[0,1]$ (in the sense of \cite{Spi}, see also \cite{ManMar2}) 
to the restriction of the bulk algebra 
to the coordinates $(g,\rho,s) \in \GL_2(\Q)\times M_2(\hat\Z)\times \tilde\cP_N$, by a semigroup
${\rm Red}_N \subset \Gamma_N$ that implements a family of continued fraction algorithms
parameterized by $N$.  The boundary algebras we construct have a semi-direct product structure $\mathcal{A}_{\partial, G, \mathcal{P}_N} = \mathcal{B}_{\partial, N} \rtimes \textrm{Red}_N$ where the $\mathcal{B}_{\partial, N}$ part of the algebra is a modified $\textrm{GL}_2$-system depending on a choice of finite index subgroup $G$ of $\GL_2(\bZ)$, and the $\textrm{Red}_N$ part is a Cuntz-Krieger-Toeplitz type algebra generated by isometries $S_{N,k}$ related to the shift of the continued fraction expansion. 

\smallskip

In this construction, in the special case where $N=1$, the action of $\GL_2(\Z)$ on 
$\H^\pm$ of the first system is replaced in the second one by the action of the shift $T$
of the usual $\GL_2$-continued fraction algorithm on $[0,1]$. This
action on $[0,1]$ is equivalent to the action of $\GL_2(\Z)$ on $\bP^1(\R)$, so
it is interpreted here as a way to describe pushing the action of $\GL_2(\Z)$ on 
$\H^\pm$ to the boundary $\bP^1(\R)$. As mentioned above, we no longer
require invariance with respect to this action, and we introduce isometries $S_k$,
associated to the partial inverses of $T$, to implement the action at the level of the
algebra. Note that, by exchanging the $\GL_2(\Z)$ action with the semigroup
action implemented by the $S_k$, this action on $[0,1]$ becomes decoupled
from the partial action of $\GL_2(\Q)$, unlike what happens on the upper half plane.
In terms of the original interpretation of the $\GL_2$-system as implementing the
commensurability relation on lattices with possibly degenerate level structure, in this 
boundary setting what remains of the commensurability relation affects the level structures 
(both through the action on $M_2(\Z)$ and on the space of cosets $\tilde\cP$) but does not change 
the pseudolattice $\Z \theta+\Z \subset \R$. The reason behind this choice is the lack 
(at present) of a good theory of isogeny for noncommutative tori, unlike the notion of
isomorphism realized by the bimodules implementing Morita equivalence. 
This means that, at the level of the noncommutative
boundary of the modular curve (which should be thought of as the moduli space of noncommutative 
tori with level structure), we see the commensurability relation as a relation on level structures.
Both the semigroup and the Hecke operators simultaneously act on the 
cosets in $\tilde\cP$, with commuting actions. 

\smallskip

A more elaborate model of the boundary algebra would require developing a good
setting for nontrivial isogenies of noncommutative tori. This can in principle be done
by considering a larger class of bimodules that are not imprimitivity bimodules associated 
to Morita equivalences (for instance, the bimodules constructed  in Proposition~5.7 of \cite{LiMa}), 
and select among these the ones that correspond to a good notion of isogenies. While this
approach is certainly feasible, it is outside of the narrower scope of the present paper,
and will be considered elsewhere.  

\smallskip

The case $N=-1$ corresponds to the $\SL_2(\Z)$-continued fraction algorithm.
The other algebras in our family, for other values of $N$, do not have the same direct
interpretation in terms of the geometry of modular curves as the $N=\pm 1$ cases, 
because the semigroup ${\rm Red}_N$ of the continued fraction algorithm
sits inside the group $\Gamma_N$ but will no longer necessarily have, in general, the same 
orbit structure. The main reason to consider this entire family of algebras is because
the structure of KMS states of the resulting quantum statistical mechanical systems
becomes more transparent when viewed over this whole family with varying parameter $N$. 

\smallskip

We analyze the dynamics, partition function, and structure of the KMS states of the boundary algebras. For $N\neq 1$, the partition function has an analytic continuation with poles at $1$, $2$ and at a point $\beta_{N,c} \in (1,2)$ depending on the choice of $N$. In the case of $N=1$, there is no partition function. We show that the structure of KMS states depends on that of ${\rm Red}_N$, the Cuntz--Krieger--Toeplitz
type system generated by the isometries implementing the continued fraction algorithm, 
and also on the structure of KMS states on $\mathcal{B}_{\partial, N}$, our 
generalization of the $\GL_2$-system. For all $N\neq 1$ there is a critical
inverse temperature $\beta_{N,c}$ in the interval $(1,2)$ with the property that no KMS
exist for $\beta< \beta_{N,c}$. Above this critical inverse temperature there are as many extremal
KMS states as there are for our generalized $\GL_2$-system. In particular, for $\beta>2$
all the extremal KMS states are Gibbs states, and we exhibit a family of them parameterized,
when $N\neq -1$, by the set
$$ \GL_2(\Z) \backslash (\GL_2(\hat\Z) \times \tilde \cP_N)\times  \cD_{[0,1]\cap \Q} \, ,  $$
where $\tilde\cP_N$ is a space of cosets and $\cD_{[0,1]\cap \Q}$ is the disconnection of the
unit interval at the rationals, and when when $N=-1$ by the set
$$ \SL_2(\Z)\backslash (\GL_2(\hat\Z) \times \tilde \cP_{-1})\times  \cD_{[0,1]\cap \Q} \, .  $$
For $\beta \to \infty$ these Gibbs states converge weakly to KMS$_\infty$ states given by evaluation. 

\smallskip

In the special case $N=1$, the KMS states at finite $\beta$ 
disappear entirely, due to the fact that in this case the Cuntz--Krieger--Toeplitz part
has no KMS states, while only the ground states remain and satisfy a weaker
form of the KMS condition. The $\SL_2$-case with $N=-1$ is, in this respect, the
nicer in this family of algebras, as it has both the geometric
interpretation in terms of modular curves and a nicer structure of KMS states with
a convergent partition function in the low-temperature range $\beta>2$ and
Gibbs states converging to the ground states as the temperature goes to zero. 

\smallskip

The ground states are the only ones that we need in order to investigate the
pairing with limiting modular symbols. So for that purpose we can restrict to the case $N=1$,
which more closely reflects the setting of \cite{ManMar}. We introduce a class of ``boundary
arithmetic elements". These are obtained by first constructing a ``boundary value map" which
is a linear map from the bulk to the boundary algebra associated to the choice of a cusp form.
The same map can be applied to the arithmetic algebra of the bulk system (which as in the
original $\GL_2$-case is an algebra of unbounded multipliers consisting of modular functions
and Hecke operators). The subalgebra generated by the images is the arithmetic algebra
of the boundary system. The image of the boundary value map consists of elements obtained 
by a procedure of averaging along geodesics. The evaluation of the ground states on these
boundary values therefore agrees with the pairing of cusp forms with limiting modular symbols. 
This construction reflects the fact that, while the abelian class field theory of imaginary quadratic
fields arises from evaluation of modular functions at complex multiplication
points in the upper half plane, the corresponding geometry of real multiplication is expected to
depend on a suitable averaging along geodesics with endpoints at conjugate quadratic 
irrationalities in a real quadratic field. 

\smallskip

The table below summarizes the properties of the Bost-Connes system, $\GL_2$-system, and 
boundary-$\GL_2$-system. 

\begin{figure}[H]
\renewcommand{\arraystretch}{1.5}
\begin{tabular}{|p{3cm}|c|c|c|}
\hline
& Bost-Connes  & $\GL_2$ & $N$-Boundary-$\GL_2$ \\
\hline
\small{\parbox[t]{2cm}{Partition \\ function\\ }}& \small{$\zeta(\beta)$ }& \small{$\zeta(\beta) \zeta(\beta-1)$} & $\frac{\zeta(\beta)\zeta(\beta-1) \prod\limits_{ p \mid N} \left(1 - p^{-\beta} \right) \left(1 - p^{-(\beta-1)}\right)}{1 + \sum\limits_{n=1}^{N-1} n^{-\beta} -\zeta(\beta) }$ \\
\hline
\small{\parbox[t]{3cm}{Critical temps \\}} & \small{$\beta =1$ }& \small{$\beta =1,2$ }& \small{ $\beta = 1, \beta_{N,c}, 2 $}\\
\hline
\small{\parbox[t]{2.2cm}{Low-temp. \\ KMS states \\}}& \parbox[t]{1.9cm}{\centering\small{  $\GL_1(\hat\Z)$ } }
& \small{$\Gamma_{-1} \backslash( 
\GL_2(\hat{\bZ})\times \bH) $ } & 
\parbox[t]{4.7cm}{\centering
\small{ $ \Gamma_{\pm 1} \backslash ( \GL_2( \hat\Z ) \times \tilde\cP_N ) \times \cD_{[0,1]\cap \Q}$  }
}  \\ 
\hline
\parbox[t]{3.5cm}{\small{Ground states on \\ arith. algebra\\}} & \parbox[t]{2cm}{\centering \small{generators of $\bQ^{cycl}$}} & \parbox[t]{2.8cm}{\centering \small{generators of modular field $F$}}& \parbox[t]{4.8cm}{\centering \small{pairing of limiting modular symbols and cusp forms}}  \\
\hline
\small{ Symmetries}&\parbox[c]{2.2cm}{\centering \small{$\mathbb{A}_f^*/\bQ^* \simeq {\rm Gal}(\bQ^{cycl}/\bQ)$} }&\parbox[c]{3cm}{\centering \small{$\GL_2(\mathbb{A}_f)/\bQ^* \simeq {\rm Aut}(F)$} } & ? \\
\hline
\end{tabular}
\caption{Quantum statistical mechanical properties of the Bost-Connes, $\GL_2$, and Boundary-$\GL_2$ systems}
\end{figure}

\section{The $\GL_2$-system}\label{GL2Sec}

In this section we recall the basic properties of the $\GL_2$-quantum statistical mechanical 
system of \cite{CoMa}, in a version that accounts for the choice of a finite index subgroup
of $\GL_2(\Z)$ and for a more general class of subgroups of $\GL_2(\Q)$ that include
$\GL_2(\Z)$ and $\SL_2(\Z)$ as special cases.

\subsection{The $\GL_2$-quantum statistical mechanical system}

The $\GL_2$-quantum statistical mechanical system constructed in \cite{CoMa} as a generalization
of the Bost--Connes system of \cite{BC} has algebra of observables given by the non-commutative
$C^*$-algebra describing the ``bad quotient" of the space of $2$-dimensional $\Q$-lattices up
to scaling by the equivalence relation of commensurability. This algebra is made dynamical by a
time evolution defined by the determinant of the $\GL_2^+(\Q)$ matrix that implements 
commensurability. There is an arithmetic algebra of unbounded multipliers on the $C^*$-algebra
of observable, which is built in a natural way out of modular functions and Hecke operators
(see \cite{CoMa} and Chapter~3 of \cite{CoMa-book}). The KMS-states for the time evolution
have an action of $\Q^*\backslash \GL_2(\A_{\Q,f})$ by symmetries, which include both
automorphisms and endomorphisms of the $C^*$-dynamical system. The KMS states at
zero temperature, defined as weak limits of KMS-states at positive temperature, are evaluations
of modular functions at points in the upper half plane and the induced action of symmetries
on KMS-states recovers the Galois action of the automorphisms of the modular field. 
In the case of imaginary quadratic fields, the associated Bost--Connes system, constructed
in \cite{HaPa} and \cite{LLN}, can be seen as a specialization of the 
$\GL_2$-quantum statistical mechanical system of \cite{CoMa} at $2$-dimensional $\Q$-lattices that
are $1$-dimensional $\K$-lattices, with $\K$ the imaginary quadratic field, and to CM points in the
upper half plane. 

\smallskip

Our goal here is to adapt this construction to obtain a specialization of the $\GL_2$ system
to the boundary $\bP^1(\R)$ and a further specialization for real quadratic fields.  Our starting
point will be the same algebra of the $\GL_2$ system of \cite{CoMa}, hence we start by
reviewing briefly that construction in order to use it in our setting.
It is convenient, for the setting we consider below, to extend the construction 
of the $\GL_2$ system recalled above to the case where we replace $\GL_2^+(\Q)$
acting on the upper half plane $\H$ with $\GL_2(\Q)$ acting on $\H^\pm$ (the upper and
lower half planes) and we consider a fixed finite index subgroup $G$ of $\GL_2(\Z)$,
where the latter replaces $\Gamma=\SL_2(\Z) =\GL_2^+(\Q)\cap \GL_2(\hat\Z)$ in
the construction of the $\GL_2$-system. We can formulate the resulting quantum
statistical mechanical system in the following way. 
We can consider two slightly different versions of the convolution algebra.

\smallskip

\begin{defn}\label{AcG}
The involutive algebra $\cA^c_G$ is given by complex valued functions on
\begin{equation}\label{Upm}
 \cU^\pm =\{ (g,\rho,z) \in \GL_2(\Q) \times M_2(\hat \Z) \times \H^\pm\,|\, g\rho \in M_2(\hat \Z) \} 
\end{equation} 
that are invariant under the action of $G\times G$ by 
$(g,\rho,z) \mapsto (\gamma_1 g \gamma_2^{-1}, \gamma_2\rho, \gamma_2 z)$, and that 
have finite support in $g\in G\backslash \GL_2(\Q)$, 
depend on the variable $\rho \in M_2(\hat \Z)$ through the projection onto some finite
level $p_N: M_2(\hat \Z) \to M_2(\Z/N\Z)$ and have compact support in the variable $z\in \H^\pm$.
The convolution product on $\cA^c_G$ is given by 
\begin{equation}\label{convolG}
(f_1\star f_2)(g,\rho,z) =\sum_{h \in G\backslash \GL_2(\Q)\,:\, h\rho \in M_2(\hat\Z)} f_1(g h^{-1}, h \rho, h(z)) g_2(h,\rho,z) 
\end{equation}
and the involution is $f^*(g,\rho,z)=\overline{f(g^{-1}, g\rho, g(z))}$.  The algebra $\cA^c_G$ is
endowed with a time evolution given by $\sigma_t(f)(g,\rho,z) = |\det(g)|^{it} f(g,\rho,z)$.
\end{defn}

\smallskip

\smallskip

Let $\Xi$ denote the coset space $\Xi=\GL_2(\Z)\backslash \GL_2(\Q) /\GL_2(\Z)$ and let
$\Z \Xi$ denote the free abelian group generated by the elements of $\Xi$. For
simplicity of notation we write $\Gamma =\GL_2(\Z)$. The following facts are well known
from the theory of Hecke operators. For any double coset $T_\alpha=\Gamma \alpha \Gamma$
in $\Xi$, there are finitely many $\alpha_i \in \Gamma \alpha \Gamma$ such that 
$\Gamma \alpha \Gamma=\sqcup_i \Gamma \alpha_i$. Thus, one can define a product
on $\Xi$ by setting 
\begin{equation}\label{HeckeOpsProd}
 T_\alpha T_\beta =\sum_\gamma c_{\alpha\beta}^\gamma T_\gamma 
\end{equation} 
where for $\Gamma \alpha \Gamma =\sqcup_i \Gamma \alpha_i$ and 
$\Gamma\beta \Gamma =\sqcup_j \Gamma \beta_j$, the coefficient 
$c_{\alpha\beta}^\gamma$ counts the number of pairs $(i,j)$ such that
$\Gamma \alpha_i\beta_j =\Gamma \gamma$. The ring structure on $\Z\Xi$
determined by the product \eqref{HeckeOpsProd} can equivalently be described
by considering finitely supported functions $f: \Xi \to \Z$ with the associative convolution product
\begin{equation}\label{HeckeConv}
(f_1\star f_2)(g) =\sum_h f_1(g h^{-1}) f_2(h) 
\end{equation}
where the sum is over the cosets $\Gamma h$ with $h\in \GL_2(\Q)$ or
equivalently over $\Gamma\backslash \GL_2(\Q)$.  The Hecke operators
are built in this form into the algebra of the $\GL_2$-system, through the
dependence on the variable $g\in \Gamma\backslash \GL_2(\Q)/\Gamma$,
see the discussion in \cite{CoMa-book}, Proposition~3.87.

\smallskip
\subsubsection{Coset spaces}\label{CosetSec}

We introduce here a variant $\cA^c_{\GL_2(\Z),G,\cP}$ of the $\GL_2$-algebra, where
an additional variable is introduced that accounts for the choice of the finite index
subgroup $G\subset \GL_2(\Z)$ through the coset spaces $\bP_\alpha=\GL_2(\Z) \alpha G/G$,
for $\alpha \in \GL_2(\Q)$, which include for $\alpha=1$ the coset space $\bP=\GL_2(\Z)/G$.
Let $\tilde\cP$ denote the product space $\tilde\cP=\prod_\alpha \bP_\alpha$. Consider the
$\Z$-modules $\Z\cP$, identified with finitely supported $\Z$-valued functions on $\cP$,  
and $\Z \tilde\cP$, of finite $\Z$-valued functions on $\tilde\cP$, namely functions that
factor through a projection of $\tilde\cP$ to a finite product of $\bP_\alpha$. We can identify
$\Z \tilde\cP$ with the bosonic Fock space
\begin{equation}\label{ZtildeP}
\Z \tilde\cP =\cS(\Z\cP)=\oplus_n (\Z\cP)^{\otimes n} \, . 
\end{equation}
We write $\C\cP:=\Z\cP\otimes_\Z \C$ and $\C\tilde\cP:=\Z\tilde\cP\otimes_\Z \C$.

\begin{lem}\label{cosetsHecke}
Let $G\subset \GL_2(\Z)$ be a finite index subgroup such that
$\alpha G \alpha^{-1} \cap \GL_2(\Z)$ is also a finite index subgroup, for all $\alpha \in \GL_2(\Q)$.
Consider the double coset $\GL_2(\Z) \alpha G$, with the left action of $\GL_2(\Z)$ and the right action
of $G$.  The orbit spaces $\bP_\alpha=\GL_2(\Z) \alpha G/G$ are finite. The algebra $\Z\Xi$ of Hecke 
operators acts on the modules $\Z \cP$ and $\Z\tilde\cP$.   
\end{lem}

\proof We show that the map $\GL_2(\Z)\to \GL_2(\Z) \alpha G$ given by multiplication $\gamma \mapsto \gamma \alpha$ induces a bijection between $\GL_2(\Z)/(\alpha G \alpha^{-1} \cap \GL_2(\Z))$ and $\GL_2(\Z) \alpha G/G$, hence the orbit space $\bP_\alpha=\GL_2(\Z) \alpha G/G$ is finite. 
For $\ell=\alpha g \alpha^{-1} \in \alpha G \alpha^{-1} \cap \GL_2(\Z)$ and
$\gamma \in \GL_2(\Z)$ we have $\gamma \ell \alpha  =\gamma \alpha g \sim \gamma \alpha$ in  
$\Gamma \alpha G/G$ so the map is well defined on equivalence classes. It is injective
since two $\gamma,\gamma'\in \Gamma$ with the same image differ by 
$\gamma'\gamma^{-1}=\alpha \ell \alpha^{-1}$ in $\alpha G \alpha^{-1} \cap \Gamma$ and it is also 
surjective by construction. The action of the algebra $\Z\Xi$ of Hecke operators is given by the
usual multiplication of cosets. We write $\Gamma \alpha \Gamma=\sqcup_i \Gamma \alpha_i$ for
finitely many $\alpha_i \in \Gamma \alpha \Gamma$ and $\Gamma \beta G=\sqcup_j \Gamma \beta_j$ for
finitely many $\beta_j \in \Gamma \beta G$. The product is then given by 
$$ \Gamma \alpha \Gamma \cdot \Gamma \beta G =\sum_\gamma c_{\alpha\beta}^\gamma \Gamma \gamma G$$ where $c_\gamma$ counts the number of pairs $(i,j)$ such that $\Gamma \alpha_i \beta_j =\Gamma \gamma$.
This induces an action on $\Z \cP$ by linearity and on $\Z\tilde\cP$ by multilinearity,
\begin{equation}\label{multilinact}
 \Gamma \alpha \Gamma \cdot (\Gamma \beta_1 G \otimes \cdots \otimes \Gamma \beta_n G)=
\sum_{\gamma_1, \ldots, \gamma_n} c^{\gamma_1}_{\alpha \beta_1} \cdots c^{\gamma_n}_{\alpha \beta_n}  \,\, (\Gamma \gamma_1 G \otimes \cdots \otimes \Gamma \gamma_n G) \, . 
\end{equation}
\endproof 

\smallskip

The condition that $\alpha G \alpha^{-1} \cap \GL_2(\Z)$ is also a finite index subgroup,
for all $\alpha \in \GL_2(\Q)$ is certainly satisfied, for instance, when $G$ is a congruence
subgroup.

\smallskip

In terms of generators $T_\alpha$ in the Hecke algebra $\Z\Xi$ and an element $\sum_s a_s \delta_s$
in $\Z\cP$, we write the action of $\Z\Xi$ on the module $\Z\cP$ as
\begin{equation}\label{actionT1}
T_\alpha \sum_s a_s \delta_s = \sum_s a_s T_\alpha \delta_s =\sum_s a_s \sum_i c_{\alpha,s}^i \delta_{s_i} =
\sum_i (\sum_s a_s c_{\alpha,s}^i) \delta_{s_i}.
\end{equation}
We can equivalently write elements of $\Z\cP$ as finitely
supported functions $\xi: \cP \to \Z$, and elements of $\Z\Xi$ as finitely supported functions $f: \Xi \to \Z$,
and write the action in the equivalent form
\begin{equation}\label{actionT2}
(f \star \xi)(s)=\sum_h f(gh^{-1}) \xi(hs)
\end{equation}
where the sum is over cosets
$\Gamma h$ and for $\xi(s)=\sum_\sigma a_\sigma \delta_\sigma(s)$ we write $\xi(hs)$ as
\begin{equation}\label{actionT3}
 \xi(hs):=\sum_i (\sum_\sigma a_\sigma c_{h,\sigma}^i) \delta_{s_i}(s).
\end{equation}
The action on $\Z\tilde\cP$ is then rephrased analogously, according to \eqref{multilinact}.

\smallskip
\subsubsection{More general subgroups and coset spaces}

We will also want to consider a more general family of double coset spaces in order to consider all possible $N$-continued fraction expansions as described at the beginning of Section \ref{boundary-system}. 

\begin{defn}\label{GammaNdef}
For $N\in \Z \backslash \{0\}$, let $\cG_N$ denote the subgroup of $\GL_1(\Q)$ 
generated by $-1$ and the prime factors of $N$. Let $\Delta$ be the diagonal
subgroup of $\GL_2(\Q)$,
$$ \Delta=\{ g\in \GL_2(\Q) \,|\, g=\begin{pmatrix} \lambda_1 & 0 \\ 0 & \lambda_2 \end{pmatrix}\, , \ \lambda_i \in \GL_1(\Q) \} $$
and consider the subgroups
\begin{equation}\label{DNgrp}
 \Delta_N =\{ g\in \GL_2(\Q) \,|\, g=\begin{pmatrix} \lambda_1 & 0 \\ 0 & \lambda_2 \end{pmatrix}\, , \  \lambda_i \in \cG_N  \}\, . 
\end{equation}  
For $N=1$, let $\Gamma_1 := \GL_2(\Z)$ and
for $N=-1$ let $\Gamma_{-1} := \SL_2(\Z)$, and for $|N|>1$, 
\begin{equation}\label{GammaN}
\Gamma_N = \langle \Delta_N , \GL_2(\Z) \rangle \subset \GL_2(\Q)\, ,
\end{equation}
the join of the subgroups $\Delta_N$ and $\GL_2(\Z)$, that is, the smallest
subgroup of $\GL_2(\Q)$ that contains $\GL_2(\Z)$ and the diagonal
matrices in $\Delta_N$. 
Let $\Xi_N$ denote the coset space $\Xi_N = \Gamma_N \backslash \GL_2(\Q) / \Gamma_N$,
with  $\Xi_1 = \Xi$, as before.
\end{defn}

\smallskip

\begin{lem}\label{lemGammaNGNQ}
For $N\neq -1$, let $G_{N,\Q}\subset \GL_2(\Q)$ be the subgroup
\begin{equation}\label{GammaN2}
G_{N,\Q} := \{ g \in \GL_2(\Q) | \det(g) \in \cG_N \}
\end{equation}
We have $\Gamma_N\subset G_{N,\Q}$, but the
group $\Gamma_N$ does not contain $\SL_2(\Q)$,
while $G_{N,\Q}$ does.
The group $\Gamma_N$ is generated by the
elements
\begin{equation}\label{sigmarhogens}
\sigma=\begin{pmatrix} 1 & 0 \\ 1 & 1 \end{pmatrix}\, , \ \ \ \  \rho=\begin{pmatrix} 0 & 1 \\ 1 & 1 \end{pmatrix} \, ,
\end{equation}  
\begin{equation}\label{hlambda}
\eta_\lambda =\begin{pmatrix}    \lambda & 0 \\ 0 & 1  \end{pmatrix}\, , \ \ \ \ \tilde\eta_\alpha =\begin{pmatrix}   1 & 0 \\ 0 & \alpha  \end{pmatrix} \ \ \  \lambda, \alpha\in \cG_N\, .
\end{equation}
\end{lem}

\proof The inclusions $\Gamma_N\subset G_{N,\Q}$ and $\SL_2(\Q)\subset G_{N,\Q}$ 
are evident. It is also clear that $\Gamma_N$ cannot contain $\SL_2(\Q)$ because 
$\Delta\cap \Gamma_N=\Delta_N$, hence any element $\eta_\lambda \tilde\eta_{\lambda^{-1}}$ 
in $\Delta\cap \SL_2(\Q)$, where $\lambda$ contains prime factors that do 
not divide $N$ cannot be in $\Gamma_N$.
The group $\GL_2(\Z)$ has generators $\sigma$, $\rho$ as in \eqref{sigmarhogens}
with relations $(\sigma^{-1}\rho)^2 =(\sigma^{-2}\rho^2)^6 =1$, so the
elements of \eqref{sigmarhogens} and \eqref{hlambda} generate $\Gamma_N$.
\endproof

\smallskip

We may also consider a finite index subgroup $G$ of $\Gamma_N$ and associated orbit spaces $\bP_{N, \alpha} = \Gamma_N \alpha G / G$ for $\alpha \in \GL_2(\Q)$. The discussion in Lemma \ref{cosetsHecke} remains the same, where now $c_{\alpha\beta}^\gamma$ counts the number of pairs $(i,j)$ such that
$\Gamma_N \alpha_i\beta_j =\Gamma_N \gamma$. We suppress the $N$ subscript when it is clear from context. 
\smallskip
\smallskip

To be more concrete, we illustrate here some explicit examples.

\subsubsection{The $\SL_2(\Z)$ case}

In the case of the algebra of the
$\GL_2$-system of \cite{CoMa}, for an invertible $\rho\in \GL_2(\hat\Z)$, the relevant Hecke algebra
is $\Z\Xi_{-1}=\cH(\Gamma_{-1},\cM)$ where $\Gamma_{-1}=\SL_2(\Z)$ and the subsemigroup $\cM=M_2^+(\Z)$ of
$\GL_2^+(\Q)$.

In this case (see Chapter~4 of \cite{Krieg}) $\cH(\Gamma_{-1},\cM)$, as an algebra over $\Z$, 
is generated by the Hecke operators
$$ T(\ell)=\sum_{\alpha \in \Gamma_{-1} \backslash \cM(\ell)/\Gamma_{-1}} \Gamma_{-1} \alpha \Gamma_{-1} =
\sum_{ad=\ell, \, a|d} T(a,d), $$
with $\cM(\ell)=\{ \alpha \in \cM\,|\, \det(\alpha)=\ell \}$ and
$$ T(a,d)=\Gamma_{-1} \begin{pmatrix} a & 0 \\ 0 & d \end{pmatrix} \Gamma_{-1}, $$
subject to the relations, for $k,\ell\in \N$,
$$ T(\ell) T(k)=\sum_{d | \gcd\{k,\ell\}} d\, T(d,d) \, T(\frac{k\ell}{d^2}). $$
Equivalently, the Hecke algebra $\cH(\Gamma_{-1},\cM)$ splits into primary components
$$\cH(\Gamma_{-1},\cM)=\otimes_p \cH(\Gamma_{-1},\cM)_p,$$ over the set of primes $p$, where
$\cH(\Gamma_{-1},\cM)_p=\Z[T(p), T(p,p)]$.

This description of the Hecke algebra is obtained directly from the following properties
of right cosets and double cosets (Chapter~4 of \cite{Krieg}).
Given $\alpha\in \cM$, the right coset $\Gamma_{-1} \alpha$ contains a unique representative of the form
$$ \begin{pmatrix} a& b \\ 0 & d  \end{pmatrix}, \ \  \text{ with } \, a,d\in \N, \ 0\leq b < d . $$
The set $\cM(\ell)$ decomposes as a disjoint union of $\sigma_1(\ell)=\sum_{d|\ell} d$ right
$\Gamma_{-1}$-cosets, with a set of representatives given by the matrices
$$ \begin{pmatrix} a& b \\ 0 & d  \end{pmatrix}, \ \  \text{ with } \, d\in \N, \,\, 0\leq b < d, \,\, a=\ell/d. $$
Given $\alpha\in \cM$ there are $\gamma_1,\gamma_2\in \Gamma_{-1}$ and $a,d\in \N$ with $a|d$ such that
$$ \gamma_1 \alpha \gamma_2 = \begin{pmatrix} a & 0 \\ 0 & d \end{pmatrix}. $$

\subsubsection{The $\GL_2(\Z)$ case}

Here we consider also the case where $\Gamma=\GL_2(\Z)$ and $\cM=M_2(\Z)\cap \GL_2(\Q)$
is the subsemigroup of $\GL_2(\Q)$. In this case the explicit description of
$\Z\Xi=\cH(\Gamma,\cM)$ is similar to the previous case (see Chapter~5 of \cite{Krieg}), and
$\cH(\Gamma,\cM)$ is generated by the Hecke operators
$$ T(\ell)=\sum_{\alpha \in \Gamma\backslash \cM(\ell)/\Gamma} \Gamma \alpha \Gamma, $$
where here $\cM(\ell)=\{ \alpha\in M_2(\Z)\,|\,\, |\det\alpha|=\ell \}$. The Hecke algebra splits into
primary components as in the previous case. We refer the reader to \cite{Krieg} for more details.

\subsubsection{The case with congruence subgroups}

In the case where we also consider a choice of a non-trivial congruence subgroup $G\subset \Gamma$,
with a Hecke algebra $\Z\Xi=\cH(\Gamma,\cM)$ as in the previous cases (see Section~2.7 of \cite{Miyake}),
we can identify the $\Z$-module $\Z\cP$ with
$$ \Z\cP = \Z [\Gamma\backslash \cM]^G, $$
with the identification induced by the injective homomorphism of $\Z$-modules
$$ \phi: \Z \cP \to \Z[\Gamma\backslash \cM], \ \ \ \phi(\Gamma \alpha G)=\sum_i \Gamma \alpha_i, $$
for $\Gamma \alpha G = \sqcup_i \Gamma \alpha_i$ a decomposition into right-cosets.
We can then write the action of $\Z\Xi$ on $\Z\cP$ described above in \eqref{actionT2}, \eqref{actionT3} 
in the form
$$ \Gamma\beta \Gamma \cdot \xi = \sum_\alpha a_\alpha\,\, \Gamma \alpha\beta_j, $$
where
$$ \xi =\sum_\alpha a_\alpha \,\, \Gamma \alpha $$
is a $G$-invariant element in $\Z[\Gamma\backslash \cM]$ and $\Gamma\beta\Gamma =\sqcup_j \Gamma \beta_j$.
The action is independent of the choice of representatives (Lemma~2.7.3 of \cite{Miyake}).

Thus, for general elements $h\in \Z\Xi$ and $\xi \in \Z\cP$ with $h=\sum_\beta b_\beta \, \Gamma \beta \Gamma$
and $\xi =\sum_\alpha a_\alpha \Gamma \alpha G$, we reformulate \eqref{actionT2}, \eqref{actionT3} as
$$ h \star \xi =\sum_{\alpha,\beta,\gamma} b_\beta \, a_\alpha \, c_{\alpha,\beta}^\gamma \Gamma \gamma G, $$
see (2.7.3) of \cite{Miyake}. This in turn determines an action on $\Z\tilde\cP$ as in \eqref{multilinact}.

\subsubsection{The bulk algebra }

We now proceed to the construction of a  ``bulk algebra" (namely,
the algebra associated to the bulk space $\H$), which includes the choice of a finite index
subgroup $G\subset \Gamma_N$.

\smallskip

\begin{lem}\label{freetildePact}
Let $G\subset \Gamma_N$ be a finite index subgroup with 
$\bP_{N,\alpha}=\Gamma_N \alpha G/G$ for $\alpha \in \GL_2(\Q)$.
The group $\Gamma_N$ acts on 
$\tilde\cP_N=\prod_\alpha \bP_{N,\alpha}$ with stabilizer given by
the scalars in $\cG_N \cap G$, hence the group 
$\bar\Gamma_N=\Gamma_N /\cG_N$ acts freely on $\tilde\cP_N$
\end{lem}

\proof Arguing as in Lemma~\ref{cosetsHecke}, we can identify the cosets
$\bP_{N, \alpha} = \Gamma_N \alpha G / G$ with the quotients 
$\Gamma_N /(\alpha G \alpha^{-1} \cap \Gamma_N)$, hence the
action of $\Gamma_N$ on $\prod_\alpha \bP_{N, \alpha}$ has stabilizer 
$\bar G_N:=\cap_{\alpha\in \GL_2(\Q)} \alpha G \alpha^{-1} \cap \Gamma_N$. 
By construction $\cap_{\alpha\in \GL_2(\Q)} \alpha G \alpha^{-1}$ is a
normal subgroup of $\GL_2(\Q)$ so it either contains $\SL_2(\Q)$ or is 
contained in the center $Z_2(\Q)$ of $\GL_2(\Q)$. As observed in
Lemma~\ref{lemGammaNGNQ}, $\Gamma_N$ does not contain 
$\SL_2(\Q)$, hence $G \subset \Gamma_N$ also does not, and
therefore $\cap_{\alpha\in \GL_2(\Q)} \alpha G \alpha^{-1} \subset Z_2(\Q)$.
We then have $\bar G_N \subset Z_2(\Q)\cap \Gamma_N=\cG_N\, {\rm id}$,
so $\bar G_N = \cG_N \cap G$.  
\endproof 

\smallskip

\begin{defn}\label{GammaNplus}
For $N\neq -1$, let $\Gamma_N^+$ denote the semigroup $\Gamma_N^+\subset \Gamma_N$
generated by $\GL_2(\Z)$ and the matrices 
$\eta_p$, $\tilde\eta_p$ as in \eqref{hlambda}, for all primes $p|N$. 
For $N=-1$ we just take $\Gamma_{-1}^+=\Gamma_{-1}=\SL_2(\Z)$.
\end{defn}

\smallskip

\begin{lem}\label{lemSrhoN}
For any given $\rho \in M_2(\hat\Z)$, 
the action of the semigroup $\Gamma_N^+$ by left multiplication on $\GL_2(\Q)$ 
preserves the set
$$ G_\rho:=\{ g \in \GL_2(\Q) \,|\, g\rho \in M_2(\hat\Z) \} \, . $$
Thus, the quotient
\begin{equation}\label{SrhoN}
\cS_{\rho,N}=\Gamma_N^+ \backslash \{ g \in \GL_2(\Q) \,|\, g\rho \in M_2(\hat\Z) \} 
\end{equation}
is well defined and, for $N\neq -1$, there is a surjective map 
$$ \cS_{\rho,1} =\GL_2\backslash \{ g \in \GL_2(\Q) \,|\, g\rho \in M_2(\hat\Z) \} \twoheadrightarrow 
\cS_{\rho,N}. $$
\end{lem}

\proof The semigroup $\Gamma_N^+$ is a subsemigroup of $\Gamma_N \cap M_2^\times(\Z)$,
with $M_2^\times(\Z)=\{ M \in M_2(\Z)\,|\, \det(M)\neq 0 \}$. Multiplication by $M_2^\times(\Z)$
preserves $M_2(\hat\Z)$, hence, for a given $\rho \in M_2(\hat\Z)$, if an element $g \in \GL_2(\Q)$
satisfies $g\rho \in M_2(\hat\Z)$, any $h\in M_2^\times(\Z)$ the element $hg$ will also satisfy
$hg\rho \in M_2(\hat\Z)$. We have a surjection $\cS_{\rho,1}\to \cS_{\rho,N}$ since
$\Gamma_N^+$ contains $\GL_2(\Z)$ for $N\neq -1$. 
\endproof

\smallskip

\begin{rem}\label{actGrho}{\rm
Consider the space
\begin{equation}\label{UpmGXi}
 \cU^\pm_{G,\cP_N} =\{ (g,\rho,z,\xi)  \in \GL_2(\Q)\times M_2(\hat \Z) \times \H^\pm \times \tilde\cP_N
  \,|\, g\rho \in M_2(\hat \Z)  \} ,
 \end{equation} 
 The group $\Gamma_N$ acts by 
 \begin{equation}\label{act1GammaN}
 \gamma_1 : (g,\rho,z,s) \mapsto (\gamma_1 g, \rho, z, s). 
 \end{equation}
 There is a partially defined action of $\Gamma_N$ by
 \begin{equation}\label{act2GammaN}
  \gamma_2 : (g,\rho,z,s) \mapsto (g \gamma_2^{-1}, \gamma_2 \rho, \gamma_2 z, \gamma_2 s) \, , 
 \end{equation} 
 which is defined for elements $\gamma$ in the set $\Gamma_N \cap G_\rho$, with
 $G_\rho=\{ g\in \GL_2(\Q)\,|, g\rho \in M_2(\Z) \}$. For all $\rho\in M_2(\hat \Z)$ the
 set $\Gamma_N \cap G_\rho$ contains the semigroup $\Gamma_N^+$.}
\end{rem}

\smallskip

\begin{defn}\label{AcGP}
The involutive algebra $\cA^c_{\Gamma_N, G,\cP_N}$ is given by complex valued functions on the space
$\cU^\pm_{G,\cP_N}$ of \eqref{UpmGXi}, 
that are invariant on the orbits  
$$(g,\rho,z,s) \mapsto (\gamma_1 g \gamma_2^{-1}, \gamma_2 z, \gamma_2 s) $$ 
with the action of $\Gamma_N$ as in \eqref{act1GammaN} and
the semigroup action of $\Gamma_N^+$ as in \eqref{act2GammaN}.
Moreover, functions in $\cA^c_{\Gamma_N, G,\cP_N}$  have finite support in 
$\Gamma_N \backslash \GL_2(\Q)$ and are finite in $\tilde\cP_N$ (in the sense that they depend
on $s\in \tilde\cP_N$ through a finite projection, as in \S \ref{CosetSec}); 
they have compact support in $z\in \H^\pm$, and they depend on the variable $\rho \in M_2(\hat \Z)$ through the 
projection onto some finite level $p_n: M_2(\hat \Z) \to M_2(\Z/n\Z)$. 
The convolution product of $\cA^c_{\Gamma_N, G,\cP_N}$ is given by
\begin{equation}\label{AGPconvol}
 (f_1\star f_2)(g,\rho,z,s) =\sum_{h \in \cS_{\rho,N}} f_1(g h^{-1}, h \rho, h(z), hs) f_2(h,\rho,z, s),
\end{equation} 
with $\cS_{\rho,N}$ as in \eqref{SrhoN},  
where we are using the notation \eqref{actionT2}, \eqref{actionT3} for the action of Hecke operators
on functions of $\tilde \cP_N$. 
The involution is $f^*(g,\rho,z,s)=\overline{f(g^{-1}, g\rho, g(z), g s)}$. The algebra
$\cA^c_{\Gamma_N, G,\cP_N}$  is endowed with a time evolution given by $$\sigma_t(f)(g,\rho,z,s) = 
|\det(g)|^{it} f(g,\rho,z,s).$$
\end{defn} 

\smallskip

We focus here on the algebra $\cA^c_{\Gamma_N, G,\cP_N}$ and we
construct Hilbert space representations analogous to the
ones considered for the original $\GL_2$-system. 

\smallskip

Consider then 
the Hilbert space $\cH_{\rho,N} =\ell^2(\cS_{\rho,N})$, and the representations
$$\pi_{(\rho,z,s)}: \cA^c_{\Gamma_N,G,\cP_N} \to \cB(\cH_{\rho,N})$$
$$ \pi_{(\rho,z,s)}(f) \xi (g) =\sum_{h\in \cS_{\rho,N}} f(gh^{-1}, h\rho, h(z), hs) \xi(h). $$

We can complete the algebra $\cA^c_{\Gamma_N,G,\cP_N}$ to a $C^*$-algebra $\cA_{\Gamma_N,G,\cP_N}$ 
in the norm $\| f \|=\sup_{(\rho,z,s)} \| \pi_{(\rho,z,s)}(f) \|_{\cB(\cH_{\rho,N})}$.
The time evolution is implemented in the representation $\pi_{(\rho,z,s)}$ by the Hamiltonian
$H \xi (g) = \log |\det(g)| \, \xi(g)$.

\smallskip

\subsection{The arithmetic algebra}\label{arithmSec}

We proceed exactly as in the case of the $\GL_2$-system of \cite{CoMa} to
construct an arithmetic algebra associated to $\cA_{\Gamma_N,G,\cP_N}$. As in
\cite{CoMa} this will not be a subalgebra but an algebra of unbounded multipliers.

\smallskip

The arithmetic algebra $\cA^{ar}_{\Gamma_N,G,\cP_N}$ is the algebra over $\Q$ obtained as follows.
We consider functions on $\cU^\pm_{G,\cP_N}$ of \eqref{UpmGXi} that are invariant under the action of $\Gamma_N \times \Gamma_N^+$ by $(g,\rho,z,s) \mapsto (\gamma_1 g \gamma_2^{-1}, \gamma_2 \rho, \gamma_2 z, \gamma_2  s)$  as in \eqref{act1GammaN} and \eqref{act2GammaN},
and that are finitely supported in $g\in G\backslash\GL_2(\Q)$ and finite on $\tilde\cP_N$, 
that depend on $\rho$ through some finite level projection $p_n(\rho)\in M_2(\Z/n\Z)$ and that are
holomorphic in the variable $z\in \H$ and satisfy the growth condition that $| f(g,\rho,z,s) |$ 
is bounded by a polynomial in $\max\{ 1, |\Im(z)|^{-1} \}$ when $|\Im(z)|\to \infty$. The resulting algebra
$\cA^{ar}_{\Gamma_N,G,\cP_N}$ acts, via the convolution product \eqref{AGPconvol}, as unbounded
multipliers on the algebra $\cA_{\Gamma_N,G,\cP_N}$. This construction and its properties are
completely analogous to the original case of the  $\GL_2$-system and we refer the reader to
\cite{CoMa}, \cite{CoMa-book} for details.

\smallskip

The invariance property replaces the $G$-modularity property ($G$-invariant functions on $\H$)
with $\Gamma_N$-invariant functions on $\H\times \tilde\cP_N$. The constraint that
the action is also well defined on the level structures, 
mapping $\rho\in M_2(\hat\Z)$ to $M_2(\hat\Z)$, requires restriction to an 
appropriate sub-semigroup of $\Gamma_N$. These functions
are endowed with the same convolution product \eqref{AGPconvol}. 

\medskip

\section{Boundary $\GL_2$-system \label{boundary-system}}

We now consider how to extend this setting to incorporate the boundary $\bP^1(\R)$
of the upper half-plane $\H$ and $\Q$-pseudo-lattices generalizing the $\Q$-lattices 
of \cite{CoMa}. A brief discussion of the boundary compactification of
the $\GL_2$-system was given in \S 7.9 of \cite{CoMa-book} and in \cite{CMR2} and, 
but the construction
of a suitable quantum statistical mechanical system associated to the boundary
was never worked out in detail. 

\smallskip

We replace the full $\bP^1(\R)$ boundary of $\H^\pm$ with the smaller interval $[0,1]$.
In the $N=1$ case this choice is natural as this interval meets every orbit of the
$\GL_2(\Z)$ action and the equivalence relation given by this action can be 
described equivalently through the shift $T$ of the continued fraction expansion.
This action can be implemented via the semigroup of
reduced matrices in the form of a crossed product algebra. Inspired by this case,
we adopt the same setting for the whole family of algebras parameterized by the
nontrivial integer $N$, with corresponding continued fraction algorithms on the
interval $[0,1]$ and associated semigroups. 
We analyze Hilbert space representations, time evolution, Hamiltonian,
partition function and KMS states.

\smallskip
\subsection{Continued fraction algorithms}\label{contfrSec}

We consider the countable family of $N$-continued fraction expansions given by 

\begin{equation}\label{cntdfraction}
[a_0; a_1, a_2, a_3, ... ]_N = a_0 + \frac{N}{a_1 + \frac{N}{a_2 + \frac{N}{a_3 + ...}}}
\end{equation}
with $a_i \geq N$ when $N\geq 1$ and $a_i \geq |N| +1$ when $N \leq -1$. We denote the set of allowed digits of the $N$-continued fraction expansion by $\Phi_N$, 
\begin{equation}\label{PhiNeq}
 \Phi_N = \begin{cases} \N_{\geq N}& \textrm{when } N \geq 1 \\
 \N_{\geq |N| +1} &\textrm{when } N \leq-1 \end{cases}.
 \end{equation}
 where we write $\N_{\geq N}:=\{ n\in \N \,|\, n\geq N \}$. 

For each $N$-continued fraction expansion, we introduce an algebra associated to
the boundary $\bP^1(\R)$ with the action of a certain subsemigroup of $\GL_2(\Q)$, called the semigroup of reduced matrices, depending on the choice of $N$. In the case with $N=1$ this semigroup of reduced matrices is contained in $\GL_2(\Z)$ and in the case with $N=-1$ it is contained in $\PSL_2(\bZ)$. In the $N = \pm 1$ cases, the associated algebra can be interpreted as a boundary algebra of the $\GL_2$ system. While we have no similar direct geometric interpretation when $|N|>1$, considering the whole family of systems leads to some interesting observations about the structure of the KMS states. 

\smallskip
\subsection{Boundary dynamics and coset spaces}
 
The $N$-continued fraction expansion of a real number $x$ can be retrieved via the shift operator $T_N : [0,1] \rightarrow [0,1]$ given by

\begin{equation}\label{shift-def}
T_N(x) = \frac{N}{x} - \left\lfloor \frac{N}{x} \right\rfloor , \;\; x \neq 0 ;  \;\;\;\; T_N(0) = 0. 
\end{equation}

For $x\in[0,1)$, one has that $a_0 = 0$ and $a_i = \left\lfloor \frac{N}{T_N^{i-1}(x)} \right\rfloor$ in the case that $N\geq 1$, and $a_0 = 1$ and $a_i = - \left\lfloor \frac{N}{T_{N}^{i-1}(1-x)} \right\rfloor$ in the case that $N \leq -1$. 

We extend $T_N$ to a map on $ [0,1] \times \bP$ by 
\begin{equation}\label{shift-def-2}
T_N : (x,s) \mapsto \left( \frac{N}{x}-\left\lfloor \frac{N}{x} \right\rfloor, \begin{pmatrix} -\lfloor N/x\rfloor & N \\ 1 & 0 \end{pmatrix} \cdot s \right). 
\end{equation} 
We remark that in the geometrically meaningful case of $N=1$, the set $[0,1]\times \bP$ meets every orbit of the action of $\GL_2(\Z)$ on $\bP^1(\R)\times \bP$,
acting on $\bP^1(\R)$ by fractional linear transformations and on $\bP= \GL_2(\Z)/G$ by the left-action
of $\GL_2(\Z)$ on itself. Moreover, two points $(x,s)$ and $(y,t)$ in $[0,1]\times \bP$ are in the same
$\GL_2(\Z)$-orbit iff there are integers $n,m\in \N$ such that $T_1^n(x,s)=T_1^m(y,t)$.

\smallskip

\begin{lem}\label{actionP}
The action of the shift map \eqref{shift-def-2} on $[0,1]\times \bP$ extends to an action on $[0,1]\times \bP_{N,\alpha}$,
with $\bP_{N,\alpha}=\Gamma_N \alpha G/G$, for any given $\alpha \in \GL_2(\Q)$,
hence to an action on $[0,1]\times \cP_N$ and on $[0,1]\times \tilde\cP_N$ with 
$\cP_N=\cup_\alpha \bP_{N,\alpha}$ and $\tilde\cP_N=\prod_\alpha \bP_{N,\alpha}$.
\end{lem}

\proof The action of $T_{N}$ on $(x,s)\in [0,1]\times \bP$ is implemented by the action of the matrix
$$ \begin{pmatrix}  - \lfloor N/x \rfloor & N \\ 1 & 0 \end{pmatrix} \in \Gamma_N. $$
The same matrix acts by left multiplication on $\bP_{N,\alpha}=\Gamma_N \alpha G/G$, hence
it determines a map $T_N: [0,1]\times \bP_{N,\alpha} \to [0,1]\times \bP_{N,\alpha}$.
\endproof

\subsection{Disconnection algebra}

We recall here from \cite{Spi} (see also \cite{ManMar2}) the construction of the disconnection algebra of $\bP^1(\R)$ along
$\bP^1(\Q)$ and its restriction to $[0,1]$.

\smallskip

Given a subset $B\subset\bP^1(\R)$ one considers the abelian
$C^*$-algebra $\cA_B$ generated by the algebra $C(\bP^1(\R))$ and the
characteristic functions of the positively oriented intervals 
with endpoints in $B$. If the set $U$ is dense in $\bP^1(\R)$ then the algebra obtained
in this way can be identified with the norm closure of the $*$-algebra generated 
by these characteristic functions. By the Gelfand--Naimark correspondence, the
$C^*$-algebra $\cA_B$ is the algebra of continuous functions on a compact Hausdorff topological space,
$\cA_B \simeq C(\cD_B)$. We refer to this space $\cD_B$ as the {\em disconnection of
$\bP^1(\R)$ along $B$}. The space $\cD_B$ is totally disconnected iff $B$ is dense in $\bP^1(\R)$. 

\smallskip

In particular, the disconnection $\cD_{\bP^1(\Q)}$ of $\bP^1(\R)$ along $\bP^1(\Q)$ can be
identified with the ends of the tree of $\PSL_2(\Z)$ embedded in the hyperbolic plane $\H$, see
the discussion in \S 5 of \cite{ManMar2}. 

\smallskip

In our setting, since the Gauss map of the continued fraction algorithms we are considering has discontinuities,
which occur at rational points, we need to work with an algebra of continuous functions over a disconnection of 
the interval $[0,1]$ at the rationals. The algebra $C(\cD_{[0,1]\cap \Q})$ of the 
disconnection $\cD_{[0,1]\cap \Q}$ of $[0,1]$ along the rational points $[0,1]\cap \Q$ is the image
of $C(\cD_{\bP^1(\Q)})$ under the projection given by the characteristic function $\chi_{[0,1]}$ of the
interval, which is a continuous function in $C(\cD_{\bP^1(\Q)})$ by construction. 

\smallskip

\begin{lem}\label{gNkactf}
The action
\begin{equation}\label{fgNkcat}
 f \mapsto \chi_{X_{N,k}} \cdot f\circ g_{N,k}^{-1}  \ \ \ \text{ and } \ \ \  \tilde f\mapsto  f\circ g_{N,k}\, , 
\end{equation} 
with 
\begin{equation}\label{gkmat}
g_{N,k} =\begin{pmatrix}  0 & N \\ 1 & k \end{pmatrix} \ \ \  \text{ and } \ \ \  g_{N,k}^{-1} = \begin{pmatrix} -\frac{k}{N} & 1 \\ \frac{1}{N} & 0 
\end{pmatrix} \, ,
\end{equation}
is well defined on $C(\cD_{[0,1]\cap \Q})$.
\end{lem}

\proof This is immediate from \eqref{PhiNeq}, \eqref{shift-def}, \eqref{shift-def-2} but for the
convenient of the reader, we spell it out in full. Indeed, for $x\in [0,1]$, we have
$$  g_{N,k}(x) = \frac{N}{x+k} \in [0,1] $$
since $k\geq N$ by \eqref{PhiNeq}, so $f\circ g_{N,k}$ is still a function in $C(\cD_{[0,1]\cap \Q})$, while for
$x\in X_{N,k}$ we have
$$ g_{N,k}^{-1}(x) = \frac{-kx+N}{x} \in [0,1]\, , $$
because $X_{N,k}$ is the set of those $x$ for which $k=\lfloor \frac{N}{x} \rfloor$, so that $k\leq N/x \leq k+1$. Thus,
even though $f\circ g_{N,k}^{-1}$ is not necessarily in $C(\cD_{[0,1]\cap \Q})$ the product $\chi_{X_{N,k}} \cdot f\circ g_{N,k}^{-1}$
is in $C(\cD_{[0,1]\cap \Q})$.
\endproof

\subsubsection{Disconnection algebra and coset spaces}

We incorporate the coset spaces in the construction of the disconnection algebra in the following way.

\smallskip 
\begin{lem}\label{actionscommute}  
Let $\C\tilde\cP_N=\Z\tilde\cP_N\otimes_\Z \C$, with $\Z\tilde\cP_N$ as in \eqref{ZtildeP}, and
let $$\cB_N=C(\cD_{[0,1]\cap\Q},\C\tilde\cP_N)$$ denote the algebra of continuous 
functions from the disconnection of $[0,1]$ at the rationals to $\C\tilde\cP_N$.
Consider the action of the semigroup $\Z_+$ on $\cB_N$ determined by 
the action of $T_N$ on $[0,1]\times \tilde\cP_N$ of Lemma~\ref{actionP} and the action on $\cB_N$
by Hecke operators acting on $\C\tilde\cP_N$. These two actions commute.
\end{lem}

\proof We write functions $f \in \cB_N$ in the form $\sum_\alpha f_\alpha(x,s_\alpha) \delta_\alpha$
where $\delta_\alpha$ is the characteristic function of $\P_{N,\alpha}=\Gamma_N \alpha G/G$ and
$s_\alpha \in \P_{N,\alpha}$. The action of $\Z_+$ is given by 
$$ T_N^n: \sum_\alpha f_\alpha(x,s_\alpha)\delta_\alpha \mapsto \sum_\alpha f_\alpha(T_N^n(x,s_\alpha))\delta_\alpha, $$
with $T_N(x,s_\alpha)$ as in \eqref{shift-def-2}, while the action of a Hecke operator $T_\beta$ is given by
\begin{equation}\label{THeckeact}
 T_\beta: \sum_\alpha f_\alpha(x,s_\alpha)\delta_\alpha \mapsto  \sum_\gamma (\sum_\alpha c_{\beta,\alpha}^\gamma f_\alpha(x, s_\alpha)) \delta_\gamma, 
\end{equation} 
with $c_{\beta\alpha}^\gamma$ defined as in \eqref{actionT1}, modified appropriately for the choice of $N$. It is then clear that these two actions commute. 
\endproof

\smallskip

\begin{lem}\label{Tactiongk}
Let $X_{N,k} \subset [0,1]$ be the subset of points $x\in [0,1]$ with $N$-continued fraction expansion
 starting with the digit $k\in \Phi_N$, as in \eqref{PhiNeq}. 
 Let $\cB_N$ denote the algebra of continuous complex valued
functions on $\cD_{[0,1] \cap\Q}\times \tilde\cP_N$. Let $\tau_N(f)=f\circ T_N$ denote the action of the 
shift $T_N: [0,1] \to [0,1]$ of the $N$-continued fraction expansion on $f\in \cB_N$. 

Let $\cB_N$ act as multiplication
operators on $L^2([0,1],d\mu_{N})$ with $d\mu_{N}$ the $T_{N}$-invariant measures on $[0,1]$,
$$ d\mu_N(x)= \begin{cases}
\left(\log \frac{N+1}{N}\right)^{-1}  \left(N+x\right)^{-1} \, dx &\mbox{if } N \in \Z \backslash \{0,-1\}\\
\left(1-x\right)^{-1}dx &\mbox{if } N =-1
 \end{cases} .$$
With the notation \eqref{gkmat}, consider the
operators 
\begin{equation}\label{Skops}
S_{N,k} \xi (x) =\chi_{X_{N,k}}(x) \cdot \xi (g_{N,k}^{-1} x) \,  \ \ \ \text{ and } \ \ \  \tilde S_{N,k} \xi (x) = \xi( g_{N,k}\, x ),
\end{equation}

for $\xi\in L^2([0,1],d\mu_{N})$, with $\chi_{X_{N,k}}$ the characteristic function of the subset $X_{N,k} \subset [0,1]$.

These satisfy $\tilde S_{N,k}=S_{N,k}^*$ with $S_{N,k}^* S_{N,k} =1$ and $\sum_k S_{N,k} S_{N,k}^*=1$.
They also satisfy the relation 
\begin{equation}\label{TSk}
\sum_k S_{N,k} \, f \, S_{N,k}^* =f\circ T_N .
\end{equation}
\end{lem}

\proof
The shift map of the continued fraction expansion, given by $T_N(x) =N/x - [N/x]$, 
acts on $x\in X_{N,k}$ as $x\mapsto g_{N,k}^{-1} x$ with the matrix $g_{N,k}^{-1}$ acting by fractional linear transformations.
The operators $S_{N,k}$ defined as in \eqref{Skops} are not isometries
on $L^2([0,1],dx)$ with respect to the Lebesgue measure $dx$. However, if we consider the
 $T_N$-invariant probability measures $d\mu_N$, then we have
$d\mu_N\circ g_{N,k}^{-1}|_{X_{N,k}}=d\mu |_{X_{N,k}}$ for all $k\in \N$, hence 
\begin{align*}
\langle S_{N,k} \xi_1, S_{N,k} \xi_2\rangle&=\int_{X_{N,k}} \bar \xi_1\circ g_{N,k}^{-1}\, \xi_2\circ g_{N,k}^{-1} \, d\mu_N \\
&=\int_{X_{N,k}} \bar \xi_1\circ g_{N,k}^{-1}\, \xi_2\circ g_{N,k}^{-1} \, d\mu_N\circ g_{N,k}^{-1} =\int_{[0,1]} \bar \xi_1\, \xi_2\, d\mu_N 
=
\langle \xi_1, \xi_2\rangle.
\end{align*}

 We have $\tilde S_{N,k}\, S_{N,k} \xi(x) = \xi(x) \chi_{X_{N,k}}(g_{N,k} x)= \xi(x)$.
Moreover, $\tilde S_{N,k}=S_{N,k}^*$ in this inner product since we have
\begin{align*}
\langle \xi_1, S_{N,k} \xi_2\rangle &=\int_{X_{N,k}} \bar \xi_1\, \xi_2\circ g_{N,k}^{-1} \, d\mu_N \\
&= \int_{X_{N,k}} \bar \xi_1\, \xi_2\circ g_{N,k}^{-1} \, d\mu_N\circ g_{N,k}^{-1} = \int_{[0,1]} \bar \xi_1\circ g_{N,k}\, \xi_2 \, d\mu_N =
\langle \tilde S_{N,k} \xi_1, \xi_2 \rangle.
\end{align*}
Using Lemma~\ref{gNkactf}, we also have $\sum_{k} S_{N,k} \, f \, S_{N,k}^* \xi(x) = \sum_{k} f(g_{N,k}^{-1} x)  \chi_{X_{N,k}}(x) \xi(x) =f(T_N(x))  \xi(x)$.
Thus we obtain $\sum_k S_{N,k} \, f \, S_{N,k}^* =f\circ T_N$, which in particular also implies 
$\sum_k S_{N,k} \, S_{N,k}^* =1$. 
\endproof

\smallskip
\subsection{Semigroups}

Consider the set of matrices in $\GL_2(\Q)$ 
\begin{equation}\label{Redn}
{\rm Red}_{N,n}:=
\begin{cases}
\left\{ \,
\begin{pmatrix}
0 & N\\
1 & k_1
\end{pmatrix} \ \dots \
\begin{pmatrix}
0 & N\\
1 & k_n
\end{pmatrix}\, \bigg| k_i\in\Z_{\ge N} \right\}  

&\mbox{if } N \ge 1\\
\\

\left\{ \,
\begin{pmatrix}
0 & N\\
1 & k_1
\end{pmatrix} \ \dots \
\begin{pmatrix}
0 & N\\
1 & k_n
\end{pmatrix}\, \bigg| k_i\in\Z_{\ge |N| +1}\right\} 
&\mbox{if } N \le -1

\end{cases}
\end{equation}

Note that ${\rm Red}_{N,n} \subset \Gamma_N$ and in particular when $N=1$, ${\rm Red}_{1,n} \subset \GL_2(\Z)$, and when $N=-1$, ${\rm Red}_{-1,n} \subset \SL_2(\Z)$. 

The semigroups of reduced matrices are  defined as
\begin{equation}\label{Red}
 {\rm Red_N}:=\cup_{n\ge 1}\, {\rm Red}_{N,n}.
\end{equation} 
An equivalent description of the ${\rm Red_{1}}$ semigroup is given by (\cite{LewZag})
$$ {\rm Red_1}=\{ \begin{pmatrix} a & b \\ c & d \end{pmatrix} \in \GL_2(\Z) \,|\,  0\leq a \leq b, \, 0\leq c\leq d \}. $$

\smallskip

\begin{lem}\label{homRed}
Assigning to a matrix
\begin{equation}\label{prodRed}
 \gamma = \begin{pmatrix} 0 & N \\ 1 & n_1 \end{pmatrix} \cdots \begin{pmatrix} 0 & N \\ 1 & n_k \end{pmatrix} 
\end{equation} 
in ${\rm Red_{N}}$ the product $n_1\cdots n_k \in \N$ is a well defined semigroup homomorphism.
\end{lem}

\proof We only need to check that the representation of a matrix $\gamma$ in ${\rm Red_N}$ as
a product \eqref{prodRed} is unique so that the map is well defined. It is then by construction a
semigroup homomorphism. 

First we consider the $N=1$ case. The group $\GL_2(\Z)$ has generators 
$$ \sigma=\begin{pmatrix} 1 & 0 \\ 1 & 1 \end{pmatrix} \ \ \ \  \rho=\begin{pmatrix} 0 & 1 \\ 1 & 1 \end{pmatrix} $$
with relations $(\sigma^{-1}\rho)^2 =(\sigma^{-2}\rho^2)^6 =1$. The semigroup ${\rm Red}_1$ can be
equivalently described as the subsemigroup of the semigroup generated by $\sigma$ and $\rho$
made of all the words in $\sigma$, $\rho$ that end in $\rho$, so 
elements are products of matrices of the form $\sigma^{n-1}\rho =\begin{pmatrix} 0 & 1 \\ 1 & n \end{pmatrix}$. We have
$$ \gamma = \begin{pmatrix} 0 & 1 \\ 1 & n_1 \end{pmatrix} \cdots \begin{pmatrix} 0 & 1 \\ 1 & n_{\ell(\gamma)} \end{pmatrix} $$
where $\ell(\gamma)$ is the number of $\rho$'s in the word in $\sigma$ and $\rho$
representing $\gamma$.  The semigroup generated by $\sigma$ and $\rho$ is a free semigroup, 
as the only relations in $\GL_2(\Z)$ between these generators involve the inverse $\sigma^{-1}$. 
If an element $\gamma \in {\rm Red_1}$ had two different representations \eqref{prodRed}, for two
different ordered sets $\{ n_1, \ldots, n_k \}$ and $\{ m_1, \ldots, m_l \}$ then we would have a
relation $$\sigma^{n_1-1}\rho \sigma^{n_2-1}\rho \cdots \sigma^{n_k-1}\rho =
\sigma^{m_1-1}\rho \sigma^{m_2-1}\rho\cdots \sigma^{m_l-1}\rho$$ involving the generators
$\sigma$ and $\rho$ but not their inverses, which would contradict the fact that $\sigma$ and
$\rho$ generate a free semigroup.

Next we consider the case $N \in \Z \backslash \{-1,0,1\}$. We observe that we can decompose elements of ${\rm Red_{N}}$ in terms of $\rho$, $\sigma$ and  
\begin{equation}\label{etaN}
 \eta_N = \begin{pmatrix}
N & 0\\
0 & 1
\end{pmatrix},  
\end{equation}
a diagonal matrix depending on N since 

\begin{align*}
\begin{pmatrix}
0 & N\\
1 & n
\end{pmatrix}
= 
\begin{pmatrix}
N & 0\\
0 & 1
\end{pmatrix}
\begin{pmatrix}
0 & 1\\
1 & n 
\end{pmatrix}
= 
\eta_N\sigma^{n-1}\rho. 
\end{align*}

If an element $\gamma \in {\rm Red_N}$ for $|N|>1$ had two different representations \eqref{prodRed}, for two
different ordered sets $\{ n_1, \ldots, n_k \}$ and $\{ m_1, \ldots, m_l \}$ then we would have a
relation

\[ \eta_N\sigma^{n_1-1}\rho \eta_N \sigma^{n_2-1}\rho \cdots \eta_N \sigma^{n_k-1}\rho =
\eta_N \sigma^{m_1-1}\rho \eta_N \sigma^{m_2-1}\rho\cdots \eta_N \sigma^{m_l-1}\rho. \]
As before, there are no relations between $\rho$ and $\sigma$. There cannot be a relation involving $\eta_N$ and $\rho$ and $\sigma$. If we had $\textrm{word}(\eta_N, \rho, \sigma) = 1$ then the determinant of the left-hand side would be $\pm N^r$ where $r$ is the number of times $\eta_N$ appears in the word, while the determinant of the right-hand side would be $1$. Since we are in the case $|N|>1$, this is a contradiction.

Finally we consider the $N=-1$ case. $\PSL_2(\Z)$ can be written as a free product of cyclic groups 
\[ \PSL_2(\Z) \simeq C_2 \star C_3\]
with generators 
\begin{align}
B = \begin{pmatrix}
0 & -1\\
1 & 0
\end{pmatrix}
\;\;\;\; \textrm{and} \;\;\;\;
C = \begin{pmatrix}
1 & -1\\
1 & 0
\end{pmatrix}  
\end{align} 
of degree 2 and 3 respectively ($B^2=1$ and $C^3 = 1$). 
We can write a matrix in $\textrm{Red}_{-1}$ 
\begin{align*}
 \gamma = \begin{pmatrix} 0 & -1 \\ 1 & n_1 \end{pmatrix} \cdots \begin{pmatrix} 0 & -1 \\ 1 & n_{\ell(\gamma)} \end{pmatrix}
\end{align*}
 in terms of these generators by noting that in $\PSL_2(\Z)$, 
\begin{align*}
\begin{pmatrix}
0 & -1\\
1 & n
\end{pmatrix}  = B (CB^{-1})^n 
\end{align*}
and hence 
\begin{align} \label{psl-unique}
\gamma & = B (CB^{-1})^{n_1} \cdots B (CB^{-1})^{n_{\ell(\gamma)}} \notag \\
& = B (CB^{-1})^{n_1 -1} C^2B^{-1} (CB^{-1})^{n_2-1}C^2B^{-1} \dots C^2B^{-1} (CB^{-1})^{n_{\ell(\gamma)} -1} .
\end{align}
Since each $n_i \geq 2$, this is a reduced sequence of words in $C_2$ and $C_3$. Every element in a free product can be written uniquely as a reduced sequence of words. Furthermore, each element of the cyclic groups $C_2$ and $C_3$ can be written uniquely as $B^k$ or $C^k$ where $k$ is required to be either positive or negative. The form \eqref{psl-unique} is unique. If an element $\gamma \in \textrm{Red}_{-1}$ had two different expressions of the form \eqref{prodRed}, it would contradict this uniqueness. 

\endproof 

The relation between the semigroup $\textrm{Red}_N$ of \eqref{Red} and the group 
$\Gamma_N$ of \eqref{GammaN} is described as follows.

\begin{lem}\label{RedNGammaN}
For $N\neq -1$, the group $\Gamma_N \subset \GL_2(\Q)$ of \eqref{GammaN} 
satisfies $\langle {\rm Red}_N \rangle\subset \Gamma_N$, where
$\langle {\rm Red}_N \rangle$ is the smallest subgroup of $\GL_2(\Q)$ that 
contains the semigroup ${\rm Red}_N$, and can be equivalently described as
\begin{equation}\label{GammaNfromRedN}
\Gamma_N =\langle {\rm Red}_{p^k}\,|\, \,\, p|N\, , k\geq 1 \rangle,
\end{equation}
where $p$ ranges over the prime factors of $N$.
\end{lem}

\proof 
The group $\langle \textrm{Red}_N \rangle$ is the subgroup of 
$\GL_2(\Q)$ consisting of arbitrary products of elements in $\textrm{Red}_N$
and their inverses. To see that $\langle \textrm{Red}_N \rangle\subset \Gamma_N$,
we use again the fact that elements of $\textrm{Red}_N$ can be written in the form
$$ \eta_N\sigma^{n_1-1}\rho \eta_N \sigma^{n_2-1}\rho \cdots \eta_N \sigma^{n_k-1}\rho $$
for some $n_1,\ldots, n_k$, with 
$\eta_N$ as in \eqref{etaN} and $\sigma$ and $\rho$ as in \eqref{sigmarhogens}.
Since $\eta_N\in \Delta_N$ and $\sigma,\rho\in \GL_2(\Z)$ and 
$\Gamma_N=\langle \Delta_N, \GL_2(\Z) \rangle$, we obtain $\textrm{Red}_N\subset \Gamma_N$,
hence $\langle \textrm{Red}_N \rangle\subset \Gamma_N$. Similarly, we have
$\langle  \cup {\rm Red}_{p^k}  \rangle\subset \Gamma_N$, where 
we write $\langle  \cup  {\rm Red}_{p^k}  \rangle$ as short hand
notation for the right-hand-side of \eqref{GammaNfromRedN}. To show that 
$\langle  \cup  {\rm Red}_{p^k}  \rangle\supset \Gamma_N$, it suffices to show that the generators
of $\Gamma_N$ are contained in $\langle  \cup  {\rm Red}_{p^k} \rangle$.
First observe that the matrices $\eta_{p^k}$, with $k\in \Z$, 
are in $\langle   \cup {\rm Red}_{p^k} \rangle$ since we can write
$$ \eta_{p^k} = (\eta_{p^m} \sigma^{a-1}\rho) \cdot (\rho^{-1} \sigma^{1-a}\,\eta_{p^{-n}}) $$
with $k=m-n$, and $n,m\geq 1$. Thus, $\langle   \cup {\rm Red}_{p^k} \rangle$ contains ${\rm Red}_1$
hence it contains $\GL_2(\Z)=\langle {\rm Red}_1 \rangle$. Moreover,
any $\eta_\lambda$ as in \eqref{hlambda} with $\lambda \in \cG_N$ is in $\cup {\rm Red}_{p^k}$,
and the $\tilde\eta_\lambda$ with $\lambda \in \cG_N$ are in $\langle   \cup {\rm Red}_{p^k} \rangle$ since 
$$ \tilde \eta_{p^{-k}} = \begin{pmatrix} -1 & 1 \\ p^{-k} & 0 \end{pmatrix} 
\,\, \cdot \, \rho $$
with the second term in $\GL_2(\Z)$ and the first the inverse of
an element of ${\rm Red}_{p^k}$.
\endproof

\smallskip

\begin{lem}\label{lemRedHecke}
Let $\cB_N$ denote the algebra of continuous complex valued functions on $\cD_{[0,1]\cap \Q} \times \tilde\cP_N$
that are finitely supported in $\tilde\cP_N$.
The transformations $\alpha_\gamma(f)=\chi_{X_\gamma} \cdot f\circ \gamma^{-1}$ for $\gamma \in {\rm Red_N}$ define 
a semigroup action of ${\rm Red_N}$ on $\cB_N$. This action commutes with the
action of Hecke operators.
\end{lem}

\proof We check that $\alpha_\gamma(f)=\chi_{X_\gamma} \cdot f\circ \gamma^{-1}$ is 
a well defined semigroup action of ${\rm Red_N}$ on $\cB_N$. For $\gamma$ of the form 
\eqref{prodRed} we have $\alpha_\gamma=\alpha_{g_1}\cdots \alpha_{g_n}$ with the
factors $g_i = g_{N,k_i}$ as in \eqref{gkmat}, since for two matrices $\gamma,\gamma'$ in ${\rm Red_N}$ related
by $\gamma' =g_{N,k} \gamma$ for some $g_{N,k}$ as in \eqref{gkmat} we have
$\chi_{X_k} \cdot \chi_{X_{\gamma'}}\circ g_{N,k}^{-1}=\chi_{X_\gamma}$. 

The
commutation with the action of Hecke operators can be checked as in the case
of the shift $T_N$ in Lemma~\ref{actionscommute}. We write elements of the algebra
in the form $\sum_\alpha f_\alpha(x,s_\alpha) \delta_\alpha$
where $\delta_\alpha$ is the characteristic function of $\P_{N,\alpha}=\Gamma_N \alpha G/G$ and
$s_\alpha \in \P_{N,\alpha}$, with the action of Hecke operators as in \eqref{THeckeact}. 
The action of $\gamma \in {\rm Red}_N$ on the other hand is given by
$\alpha_{\gamma}\,  \sum_\alpha f_\alpha(x,s_\alpha)\delta_\alpha=\sum_\alpha \chi_{X_\gamma}(x)\, 
f(\gamma^{-1}(x,s_\alpha)) \delta_\alpha$. These actions commute, as in the case of 
Lemma~\ref{actionscommute}.
\endproof

\smallskip
\subsection{A boundary algebra}\label{boundary}

We now introduce an algebra associated to the boundary of the bulk-system.
In order to explain the reason behind our construction, consider first again the
bulk space, namely the upper-half-plane $\H$ or $\H\times \tilde\cP$ in the case where
we fix a choice of a finite index subgroup $G\subset \GL_2(\Z)$.

In the algebra of the system on the bulk space with $\Gamma=\GL_2(\Z)$, we consider functions
$f(g,\rho,z)$ that are invariant under the action of $\Gamma\times \Gamma$ mapping
$(g,\rho,z)\mapsto(\gamma_1 g \gamma_2^{-1},\gamma_2\rho,\gamma_2 z)$
(and similarly for the $\H\times \tilde\cP$ case). This same prescription cannot
be used to define a boundary algebra, since the action of $\Gamma=\GL_2(\Z)$ (or $\SL_2(\Z)$)
on the boundary $\P^1(\R)=\partial \H$ has dense orbits, hence requiring
this $\Gamma\times \Gamma$-invariance would force continuous functions to be constant.

\smallskip

One possible way around this problem would be to replace invariance
under the $\Gamma \times \Gamma$-action (in fact, invariance under the second copy of $\Gamma$,
as that is the one acting on the $z$ variable in the bulk, hence on the boundary
variable in $\P^1(\R)$) by taking an algebra given by a
crossed product with $\Gamma$.  A similar kind of boundary algebra was considered in
Section~4 of \cite{ManMar}.  Using a crossed product with $\Gamma$
would imply dealing with a boundary algebra that contains a copy of $C^*(\Gamma)$.
Invertible $\rho$'s would determine, as in the $\GL_2$-system, representations on
the Hilbert space $\cH=\ell^2(M_2^+(\Z))$ and in such representation the
algebra $C^*(\Gamma)$ generates a type ${\rm II}_1$ factor in $\cH$. This affects the
construction of KMS states for this algebra. Gibbs-type states with respect to
the trace $\Tr_\Gamma$ can be evaluated on elements in the commutant of this
factor, as discussed in Section~7 of \cite{CoMa}. However, here we do not make this
choice in the construction of the boundary algebra, and we leave this to
separate future work. This is tied up to the question mentioned in the introduction, of
developing a good theory of isogeny for noncommutative tori. 

\smallskip

The point of view we follow here on constructing a boundary algebra is
based instead on a different observation, namely on the fact that the orbits
of the action of $\Gamma=\GL_2(\Z)$ on $\P^1(\R)$ can be equivalently described
as the orbits of a discrete dynamical system $T$ acting on the interval $[0,1]$.
Thus, we will replace the crossed product by $G$ with a semigroup crossed
product that implements this equivalence relation as part of the algebra.
The reason why we prefer this approach to the crossed product by $G$ is
because the dynamical system $T$ used here is the same generalized
shift of the continuous fractions expansion used in \cite{ManMar} to
construct limiting modular symbols, and one of our main goals in this
paper is obtaining a boundary algebra that is especially suited to
relate to limiting modular symbols, hence this viewpoint is more natural here.

\smallskip

Moreover, as already discussed, this viewpoint allows us to see our
boundary algebra as one case ($N=\pm 1$ for $\Gamma=\GL_2(\Z)$ and $\SL_2(\Z)$,
respectively) of a countable family of algebras labelled by an integer $N$,
associated to a family of different continued fraction algorithms. Considering
this whole family of algebras will help us illustrate some interesting phenomena in the
structure of KMS states, even though only the $N=\pm 1$ cases have a direct
interpretation as boundary algebras of the respective bulk system and related to
the geometry of modular curves.

\smallskip

Thus, in the following we first restrict the boundary variable $\theta \in \cD_{\P^1(\Q)}$ to the
interval $[0,1]$, that is, to the disconnection $\cD_{[0,1]\cap \Q}$,  
because of the prior observation that the interval $[0,1]$ meets every $\GL_2(\Z)$-orbit.
Then we implement the action of the shift operator $T$ in the form of a semigroup crossed
product algebra. This corresponds to taking the quotient by the action of $T$ (hence by the
action of $\GL_2(\Z)$) in a noncommutative way, by considering a crossed product
algebra instead of an algebra of functions constant along the orbits. This will be a
semigroup crossed product with respect to the semigroup ${\rm Red_N}$ discussed
above, and in a form that will implement the action of the shift operator $T$ as
in Lemma~\ref{Tactiongk}.  We will work with the algebra of continuous functions
on the disconnection $\cD_{[0,1]\cap \Q}$.  In Corollary~\ref{BEsubalg} we will
further extend this disconnection space by including additional $T$-invariant subspaces.
The reason for this further extension will become clear when we consider
such boundary functions that are obtained as integration on certain
configurations of geodesics in the bulk space, see Lemma~\ref{SG2linT}.

\smallskip

Note that if we write, as before, $\Xi$ for the set of cosets $\Gamma \alpha \Gamma$
and $\cP$ for the set of cosets $\Gamma \alpha G$, we can identify the sets
$\cP\simeq \Xi \times \P$, with the finite coset space $\P=\Gamma/G$. It is convenient
to use this identification, so that, when we consider the shift operator $T$ (in the case $N=1$)
acting on $[0,1]\times \cP$, this can be viewed as the action of $T$ on $[0,1]\times \P$
as in \cite{ManMar}, with $T$ acting trivially on $\Xi$. The action on $\tilde\cP$ then
extends this action compatibly. 

\begin{defn}\label{subalg0}
Let $\bar\Gamma_N=\Gamma_N /\cG_N$, with $\cG_N=Z_2(\Q)\cap\Gamma_N$ 
and let $\bar \Gamma_N^+$ be the image of the semigroup $\Gamma_N^+$ under
the quotient map $\Gamma_N \to \bar\Gamma_N$. 
Let $\cA_{\partial,N}^c$ denote the associative algebra of continuous complex valued functions on
\begin{equation}\label{UG0}
 \cU_{\partial,G,N}=\{ (g, \rho, s) \in \GL_2(\Q) \times M_2(\hat \Z)\times \tilde\cP_N  \,|\, g\rho\in M_2(\hat \Z) \} 
\end{equation}
that are invariant with respect to the action of $\Gamma_N\times \bar \Gamma_N^+$ by
$(g,\rho,s) \mapsto (\gamma_1 g \gamma_2^{-1}, \gamma_2\rho, \gamma_2 s)$ 
as in \eqref{act1GammaN} and \eqref{act2GammaN}, and are finite
on $\tilde\cP_N$, in the sense of \S \ref{CosetSec}, and finitely supported in 
$\Gamma_N \backslash \GL_2(\Q)$, with the dependence
on $\rho$ through a finite level projection $p_n(\rho)\in M_2(\Z/n\Z)$, endowed with the
convolution product
\begin{equation}\label{convol0}
(f_1\star f_2) (g, \rho, s) = \sum_{h\in \cS_{\rho,N}} 
f_1(gh^{-1}, h\rho, hs) f_2(h,\rho,s) 
\end{equation}
with $\cS_{\rho,N}$ as in \eqref{SrhoN}, 
and with the involution $f^*(g, \rho, s) =\overline{f(g^{-1},g\rho,gs)}$. Let 
$\pi_{\rho,s}: \cA_{\partial,N}^c \to \cB(\cH_{\rho,N})$ be the representation 
$\pi_{\rho,s}(f)\xi(g)=\sum_h f(gh^{-1},h\rho,hs) \xi(h)$ for $h\in \cS_{\rho,N}$. 
Let $\cA_{\partial,N}$ denote the $C^*$-algebra
completion of $\cA_{\partial,N}^c$ with respect to $\| f \| =\sup_{(\rho,s)} \| \pi_{\rho,s}(f) \|_{\cH_{\rho,N}}$.
Let $\cB_{\partial,N} =C(\cD_{[0,1]\cap\Q},\cA_{\partial,N})$ be the algebra of continuous 
functions from $\cD_{[0,1]\cap\Q}$ to $\cA_{\partial,N}$,
with pointwise product
$$ (f_1\star f_2)(g,\rho,x,s)=\sum_h f_1(gh^{-1},h\rho,x,hs) f_2(h,\rho,x,s) $$
and involution $f^*(g,\rho,x,s)=\overline{f(g^{-1},g\rho,x,gs)}$.
\end{defn}

\smallskip

\begin{defn}\label{semicrossdef}
Let $\cA_{\partial, G,\cP_N}$ be the involutive associative algebra generated by $\cB_{\partial,N}$ and by
isometries $S_{N,k}$, with $k \in \Phi_N$. It has relations $S_{N,k}^* S_{N,k}=1$ and $\sum_{k} S_{N,k} S_{N,k}^* =1$ and relations of the form
\begin{equation}\label{relSkgk}
S_{N,k}\, f =\chi_{X_{N,k}}\, \cdot\, f\circ g_{N,k}^{-1}\, \cdot \,S_{N,k} \ \ \ \text{ and }  \ \ \  S_{N,k}^*\, f = f\circ g_{N,k} \, \cdot \, S_{N,k}^*,
\end{equation}
where $\chi_{X_{N,k}}$ is the characteristic function of the subset $X_{N,k}\subset [0,1]$ of points with
$N$-continued fraction expansion starting with $k$. The matrices $g_{N,k}, g_{N,k}^{-1}$ in $\GL_2(\Q)$ are as
in \eqref{gkmat}, with $f\circ g_{N,k}^{\pm} (g,\rho,x,s) = f(g,\rho, g_{N,k}^{\pm} (x,s) )$. They also satisfy 
the relation
\begin{equation}\label{relTSks}
\sum_{k\in \N} S_k f S_k^* = f \circ T_N 
\end{equation}
for all $f\in B_{\partial,N}$. Here, for $f=f(g,\rho,x,s)$, we have $$(f\circ T_N)(g,\rho,x,s)=f(g,\rho,T_N(x,s)),$$
with the action of $T_N$ on $[0,1]\times \tilde\cP_N$ as in Lemma~\ref{actionP}. The involution on $\cA_{\partial, G,\cP_N}$ is given by the involution on $\cB_{\partial,N}$ and by $S_{N,k}  \mapsto S_{N,k}^*$.
\end{defn}

In fact, the relation \eqref{relTSks} follows from the relations \eqref{relSkgk} as in Lemma~\ref{Tactiongk},
but we write it explicitly as it is the relation that implements the dynamical system $T_N$. Note that we
are implicitly using in the construction of the algebra the fact that the semigroup action of ${\rm Red_N}$
and the action of Hecke operators (that is built into the convolution product of $\cB_{\partial,N}$)
commute as in Lemma~\ref{lemRedHecke}. 

\smallskip
\subsection{Semigroup crossed product}

Several examples of semigroup crossed product algebras have been
considered in relation to quantum statistical mechanical systems, especially
in various generalizations of the Bost--Connes system. However, there is
no completely standard definition of semigroup crossed product algebra
in the literature. For our purposes here, the following setting suffices.

\begin{defn}\label{SemicrossDef}
Let $\cA$ be a $C^*$-algebra, and let $\cS$ be a countable semigroup
together with a semigroup homomorphism $\beta: \cS^{op} \to {\rm End}(\cA)$.
For $\ell\in \cS$, let $\beta_\ell(1)=e_\ell$ be an idempotent in $\cA$ and let $\alpha_\ell$
denote a partial inverse of $\beta_\ell$ on $e_\ell \cA e_\ell$.
The (algebraic) semigroup crossed product algebra $\cA\rtimes \cS$ is the involutive $\C$-algebra
generated by $\cA$ and elements $S_\ell, S_\ell^*$, for all $\ell\in \cS$ with the
relations
$$ S_\ell S_{\ell'} = S_{\ell \ell'}, \ \ \  S_\ell^* S_\ell  =1, \ \ \  S_\ell S_\ell^*=e_\ell, \ \ \  \sum_\ell S_\ell S_\ell^* =1, $$
$$ S_\ell \, X \, S_\ell^* = \alpha_\ell(X), \ \ \  S_\ell^* \, X \, S_\ell =\beta_\ell(X). $$
If $\pi: \cA \to \cB(\cH)$ is a representation
as bounded operators on a Hilbert space, and the $S_\ell$ act as isometries on $\cH$,
compatibly with the relations above, then semigroup crossed product $C^*$-algebra
(which will also be denoted by $\cA\rtimes \cS$)
is the $C^*$-completion in $\cB(\cH)$ of the above algebraic crossed product.
\end{defn}

\smallskip

\begin{lem}\label{crossRed}
The algebra $\cA_{\partial, G,\cP_N}$ can be identified with the semigroup crossed product
$\cB_{\partial,N} \rtimes {\rm Red_N}$ of the algebra $\cB_{\partial,N}$ of Definition~\ref{subalg0} and
the semigroup of reducible matrices, with respect to the action $\alpha: {\rm Red_N} \to {\rm Aut}(B_{\partial,N})$
by $\alpha_\gamma(f)=\chi_{X_{\gamma}} \cdot f\circ \gamma^{-1}$, where for $\gamma$ of the form 
\eqref{prodRed}, the set $X_{\gamma} \subset [0,1]$ is the cylinder set consisting of points with
$N$-continued fraction expansion starting with the sequence $n_1,\ldots, n_N$.
\end{lem}

\proof We see as in Lemma~\ref{lemRedHecke} that 
$\alpha_\gamma(f)=\chi_{X_\gamma} \cdot f\circ \gamma^{-1}$ defines a
semigroup action of ${\rm Red_N}$ on $\cB_{\partial,N}$.
The semigroup crossed product
algebra is generated by $B_{\partial,N}$ and by isometries $\mu_\gamma$ for $\gamma\in {\rm Red_N}$
satisfying $\mu_\gamma \mu_{\gamma'}=\mu_{\gamma \gamma'}$ for all $\gamma,\gamma' \in {\rm Red_N}$,
$\mu_\gamma^* \mu_\gamma=1$ for all $\gamma\in {\rm Red_N}$ and 
$\mu_\gamma\, f\, \mu_\gamma^* =\alpha_\gamma(f)$.  It suffices to consider isometries $\mu_{g_{N,k}} =:S_{N,k}$
associated to the elements $g_{N,k} \in {\rm Red_N}$ as in \eqref{gkmat} with $\mu_\gamma =S_{n_1}\cdots S_{n_N}$
for $\gamma\in {\rm Red_N}$ as in \eqref{prodRed}. We then see that the generators and relations of the
algebras $B_{\partial,N} \rtimes {\rm Red_N}$ agree with those of the algebra $\cA_{\partial, G,\cP_N}$ of
Definition~\ref{semicrossdef}.
\endproof

\smallskip

We consider the following variant of the boundary algebra introduced above, which will
be useful for the application discussed in the following section, see in particular Lemma~\ref{SG2linT}.

\begin{cor}\label{BEsubalg}
Let $E=\{ E_\alpha \}$ be a collection of subsets 
$E_\alpha \subset [0,1]$ that are invariant under the action of the shift $T_N$ of
the $N$-continued fraction expansion. We denote by $\cD_E$ the disconnection space
dual to the abelian $C^*$-algebra $C(\cD_E)$ generated by $C(\cD_{[0,1]\cap \Q})$ 
and by the characteristic functions $\chi_{E_\alpha}$. This then 
determines an algebra $\cA_{\partial, G,\cP_N,E}$ 
given by $B_{\partial,N,E}\rtimes {\rm Red_N}$ where $B_{\partial,N,E}=C(\cD_E,\cA_{\partial,N})$ is the algebra 
of continuous functions from the disconnection space $\cD_E$ to $\cA_{\partial,N}$ \
as in Definition~\ref{subalg0}. 
\end{cor}

\proof If the sets $E_\alpha$ are $T_N$-invariant then the algebra $B_{\partial,N,E}$ is
invariant under the action of the semigroup ${\rm Red_N}$ by 
$\alpha_\gamma(f)=\chi_{X_\gamma} \cdot f\circ \gamma^{-1}$, since for
$\gamma=g_{N,k_1}\cdots g_{N,k_n}$, the matrix $\gamma^{-1}$ acts on
$X_\gamma$ as the shift $T_N^n$. Thus, we can form the semigroup
crossed product algebra $B_{\partial,N,E}\rtimes {\rm Red_N}$ as in 
Lemma~\ref{crossRed}. 
\endproof

\smallskip
\subsection{Representations and time evolution}

Let $\cH_{\rho,N}$ be the same Hilbert space considered above, $\cH_{\rho,N}=\ell^2(\cS_{\rho,N})$ with
$\cS_{\rho,N}$ as in \eqref{SrhoN}. We will consider the case of an invertible $\rho\in \GL_2(\hat\Z)$. This choice is made to guarantee non-negative spectrum of the Hamiltonian of Proposition \ref{timeevolZ} and is also geometrically motivated by the $\GL_2(\Z)$ (i.e. $N=1$) setting as discussed in Section \ref{boundary}. 

\begin{lem}\label{GammaNinvrho}
When $\rho \in \GL_2(\hat\Z)$ is invertible, we have
$$ \cS_{\rho,N}=\Gamma_N^+\backslash G_\rho =  \Gamma_N^+ \backslash M_2^\times(\Z), $$
with $M_2^\times(\Z)=\{ M \in M_2(\Z)\,|\, \det(M)\neq 0 \}$. Moreover, for $\rho \in \GL_2(\hat\Z)$
and $N\neq -1$, we have $\{ \gamma \in \Gamma_N \,|\, 
\gamma \rho \in \GL_2(\hat\Z) \} =\GL_2(\Z)$.
\end{lem}

\proof
When  $\rho \in \GL_2(\hat\Z)$ is invertible, the condition $g\rho \in M_2(\hat\Z)$ means
$g\in \GL_2(\Q)\cap M_2(\hat\Z)=M_2^\times(\Z)$, 
with $M_2^\times(\Z)=\{ M \in M_2(\Z)\,|\, \det(M)\neq 0 \}$.
For $N=1$ and $\rho \in \GL_2(\hat\Z)$, in particular we have
\[\cS_{\rho,1} = \GL_2(\Z) \backslash \{ g \in \GL_2(\Q) | g\rho \in M_2(\hat\Z)\} = \GL_2(\Z)\backslash M_2^\times(\Z)\, ,\] 
and more generally $\cS_{\rho,N}=\Gamma_N^+\backslash M_2^\times(\Z)$.
For $N\neq -1$, if $\rho \in \GL_2(\hat\Z)$ and $\gamma\in \Gamma_N$ 
such that $\gamma \rho=\rho' \in \GL_2(\hat\Z)$
then 
$\gamma = \rho^{-1} \rho' \in \Gamma_N \cap \GL_2(\hat\Z) \subset \GL_2(\Q)\cap \GL_2(\hat\Z)$.
\endproof

\smallskip

\begin{lem}\label{SrhoNdet}
For $\rho\in \GL_2(\hat\Z)$, the set $\cS_{\rho,N}=\Gamma_N^+\backslash M_2^\times(\Z)$ 
is the set of matrices in $M_2^\times(\Z)$ with determinant not divisible by any prime factor 
of $N$, up to the equivalence relation defined by $\GL_2(\Z)$.
\end{lem}

\proof As in Definition~\ref{GammaNplus}, the semigroup 
$\Gamma_N^+$ is generated by $\GL_2(\Z)$ and the matrices 
$\eta_p$, $\tilde\eta_p$ as in \eqref{hlambda}, for all primes $p|N$. 
Consider a matrix $M \in M_2^\times(\Z)$. We can assume that $\det(M)>0$, that is,
$M\in M_2^+(\Z)$, as the other case can be treated similarly. A matrix $M\in M_2^+(\Z)$ can be written in the form
$$ M =\gamma_1 \cdot \begin{pmatrix} a & 0 \\ 0 & d \end{pmatrix} \cdot \gamma_2\, $$
where $a,d\in \N$ with $a|d$, and $\gamma_1, \gamma_2 \in \GL_2(\Z)$, since the
double cosets by $\GL_2(\Z)$ always have a unique representative of this form. 
If $p| \det(M)$ for some prime $p|N$, we can factor 
$$ M =\gamma_1 \cdot \eta_{p^k} \cdot \tilde\eta_{p^\ell} \cdot \begin{pmatrix} a' & 0 \\ 0 & d' \end{pmatrix} \cdot \gamma_2\, , $$
for some $k\leq \ell$, so that both $a'$ and $d'$ are not divisible by $p$. Thus, we can always
write $M$ as a product $M= A M'$ with $A\in \Gamma_N^+$ and $M'\in M_2^+(\Z)$ such
that $p\not| \det(M')$ for all $p|N$. 
\endproof

\smallskip

Let $\cW_N=\cup_{n} \cW_{N,n}$ denote the set of all finite sequences $k_1,\ldots, k_n$ with
$k_i \in \Phi_N$, including an element $\emptyset$ corresponding to the empty sequence. 
Consider the Hilbert spaces $\ell^2(\cW_N)$ and $\tilde\cH_{N,\rho} =\ell^2(\cW_N)\otimes \cH_{N,\rho}$.

\smallskip

\begin{lem}\label{lemRepCross}
The algebra $\cA_{\partial, G,\cP_N}=\cB_{\partial,N} \rtimes {\rm Red_N}$ acts on the Hilbert space
$\tilde\cH_{N,\rho}$ through the representations
\begin{equation}\label{repsCross}
\pi_{\rho,x,s}(f)\,  (\xi(g)\otimes \epsilon_{k_1,\ldots, k_n}) = \sum_{h\in \cS_{\rho,N}} 
f(g h^{-1}, h\rho, g_\gamma (x, hs)) \, \xi(h)\otimes \epsilon_{k_1,\ldots, k_n} 
\end{equation}
for $f\in B_{\partial,N}$ and with $g_\gamma=g_{N, k_1}\cdots g_{N, k_n}$, with $g_{N, k_i}$ as in \eqref{gkmat}, and 
\begin{equation}\label{repsCrossS}
\begin{split}
\pi_{\rho,x,s}(S_{N,k})\, (\xi\otimes\epsilon_{k_1,\ldots, k_n}) &= \xi\otimes \epsilon_{k, k_1,\ldots, k_n}, \\
\pi_{\rho,x,s}(S_{N,k}^*) (\xi\otimes \epsilon_{k_1,\ldots, k_n}) &= \left\{ \begin{array}{ll} \xi\otimes \epsilon_{k_2,\ldots, k_n} & k_1=k \\
0 & \text{otherwise.} \end{array}\right.
\end{split}
\end{equation}
In what follows we sometimes write $\pi_{\rho,x,s}(S_{N,k})$ as $S_{N,k}$ because the mapping of these operators does not depend on the choice of $(\rho, x,s)$. 
\end{lem}

\proof We check that \eqref{repsCross} gives a representation of the subalgebra $B_{\partial,N}$ and that
the operators $\pi_{\rho,x,s}(f)$, $S_{N,k}$, $S^*_{N,k}$ of \eqref{repsCross}  and \eqref{repsCrossS} satisfy
the relations $S_{N,k}^* S_{N,k}=1$, $\sum_{k} S_{N,k} S_{N,k}^*=1$, 
$S_{N,k} \pi_{\rho,x,s}(f)=\pi_{\rho,x,s}(\chi_{X_{N,k}} \, f\circ g_{N,k}^{-1}) S_{N,k}$ and 
$S_{N,k}^* \pi_{\rho,x,s}(f)=\pi_{\rho,x,s}(f\circ g_{N,k}) S_{N,k}^*$. For the first property it suffices to see
that $\pi_{\rho,x,s}(f_1 \star f_2)=\pi_{\rho,x,s}(f_1) \circ \pi_{\rho,x,s}(f_2)$. We have
$$ \pi_{\rho,x,s}(f_1 \star f_2) \, (\xi(g)\otimes \epsilon_{k_1,\ldots, k_n}) =\sum_{h\in \cS_{\rho,N}} 
(f_1\star f_2)(g h^{-1}, h\rho, g_\gamma (x, hs)) \, \xi(h)\otimes \epsilon_{k_1,\ldots, k_n}  = $$
$$ \sum_{h\in \cS_{\rho,N}} \sum_{\ell \in \cS_{\rho,N}} f_1(g h^{-1} \ell^{-1}, \ell h \rho, g_\gamma (x, \ell h s))
f_2(\ell, h \rho, g_\gamma (x, hs)) \xi(h)\otimes \epsilon_{k_1,\ldots, k_n}, $$
where we used Lemma~\ref{lemRedHecke}. This is then equal to
$$ \sum_{\ell \in \cS_{\rho,N}} f_1(g h^{-1} \ell^{-1}, \ell h \rho, g_\gamma (x, \ell h s)) (\pi_{\rho,x,s}(f_2) \xi) (\ell) \otimes \epsilon_{k_1,\ldots, k_n} = \pi_{\rho,x,s}(f_1) \pi_{\rho,x,s}(f_2) \xi \otimes \epsilon_{k_1,\ldots, k_n}.  $$
The relations $S_{N,k}^* S_{N,k}=1$ and $\sum_k S_{N,k} S_{N,k}^*=1$ follow directly from \eqref{repsCrossS}. For 
relations between the $S_{N,k},S_{N,k}^*$ and the $\pi_{\rho,x,s}(f)$, we have
$$ S_{N,k} \pi_{\rho,x,s}(f) \, \xi \otimes \epsilon_{k_1,\ldots, k_n} = S_{N,k}
\sum_{h\in \cS_{\rho,N}} f(g h^{-1}, h\rho, g_k^{-1} g_k g_\gamma (x, hs)) \, \xi(h)\otimes \epsilon_{k_1,\ldots, k_n} $$
$$ =  
\sum_{h\in \cS_{\rho,N}} f(g h^{-1}, h\rho, g_{N,k}^{-1} g_{N,k} g_\gamma (x, hs))\, \chi_{X_{N,k}}(g_{N,k} g_\gamma x) \, \xi(h)\otimes \epsilon_{k, k_1,\ldots, k_n}, $$
for $g_\gamma=g_{N, k_1}\cdots g_{N, k_n}$, with $\chi_{X_{N,k}}(g_{N,k} g_\gamma x)=1$, so we get
$$ \pi_{\rho,x,s}(\chi_{X_{N,k}}\cdot f\circ g_{N,k}^{-1}) \, S_{N,k} \, \xi \otimes \epsilon_{k, k_1,\ldots, k_n}. $$
The second relation is similar: we have
$$ S_{N,k}^* \pi_{\rho,x,s}(f) \, \xi \otimes \epsilon_{k_1,\ldots, k_n} = S_{N,k}^* 
\sum_{h\in \cS_{\rho,N}} f(g h^{-1}, h\rho, g_k^{-1} g_k g_\gamma (x, hs)) \, \xi(h)\otimes \epsilon_{k_1,\ldots, k_n} $$
$$ =  \sum_{h\in \cS_{\rho,N}} f(g h^{-1}, h\rho, g_{N,k} g_{\gamma'} (x, hs)) \, S_{N,k}^* \xi(h)\otimes 
\epsilon_{k_1,\ldots, k_n},  $$
with $g_{\gamma'}=g_{N, k_2}\cdots g_{N, k_n}$, so that we obtain
$$ \pi_{\rho,x,s}(f\circ g_{N,k})\, S_{N,k}^*\, \xi \otimes \epsilon_{k_1,\ldots, k_n}. $$
Thus, \eqref{repsCross}  and \eqref{repsCrossS} determine a representation of $\cA_{\partial, G,\cP_N}=\cB_{\partial ,N}\rtimes {\rm Red_N}$ by bounded operators on the Hilbert space $\tilde\cH_{\rho,N}$. 
\endproof

\smallskip

\begin{prop}\label{timeevolZ}
The transformations $\sigma_{N,t}(f)(g,\rho,x,s)=|\det(g)|^{it} f(g,\rho,x,s)$ and $\sigma_{N,t}(S_{N,k})=k^{it} S_{N,k}$
define a time evolution $\sigma_N: \R \to {\rm Aut}(\cA_{\partial, G,\cP_N})$. In the representations of Lemma~\ref{lemRepCross} on $\tilde\cH_{\rho,N}$ with $\rho\in \GL_2(\hat\Z)$ 
the time evolution is implemented by the Hamiltonian
\begin{equation}\label{Hsigmat}
H_N\, \xi(g)\otimes \epsilon_{k_1,\ldots, k_n}= \log (|\det(g)| \cdot k_1 \cdots k_n) \, 
\xi(g)\otimes \epsilon_{k_1,\ldots, k_n},
\end{equation}
with partition function
\begin{equation}\label{Zetabeta}
Z_N(\beta)= \Tr(e^{-\beta H_N}) =
\begin{dcases}
 \frac{\zeta(\beta)\zeta(\beta-1) \prod_{p \; prime \; : \;p \mid N} \left(1 - p^{-\beta} \right) \left(1 - p^{-(\beta-1)}\right)}{1 + \sum_{n=1}^{N-1} n^{-\beta} -\zeta(\beta) } & \textrm{if }N >1 \\
 \frac{\zeta(\beta)\zeta(\beta-1) \prod_{p \; prime \; : \;p \mid N} \left(1 - p^{-\beta} \right) \left(1 - p^{-(\beta-1)}\right)}{1 + \sum_{n=1}^{|N|} n^{-\beta} -\zeta(\beta) }  & \textrm{if } N \leq -1\\
\end{dcases}
\end{equation}
with $\zeta(\beta)$ the Riemann zeta function. In the $N=1$ case the operator $e^{-\beta H_1}$ is
not trace class for any $\beta >0$ hence there is no partition function. 
\end{prop}

\proof We have $$\sigma_{N,t}(f_1\star f_2)(g,\rho,x,s)=\sigma_{N,t} (\sum_h f_1(gh^{-1},h\rho,x,hs) f_2(h,\rho,x,s))= $$ $$
\sum_h |\det(gh)^{-1}|^{it} |\det(h)|^{it} f_1(gh^{-1},h\rho,x,hs) f_2(h,\rho,x,s)=\sigma_{N,t}(f_1)\star \sigma_t(f_2).$$
We also have $\sigma_{N,t}(S_{N,k}^*)=k^{-it} S_{N,k}^*$ and we see that the action of
$\sigma_{N,t}$ is compatible with the relations in $\cA_{\partial, G,\cP_N}$ and defines a $1$-parameter family
of algebra homomorphisms. By direct comparison between \eqref{Hsigmat} and \eqref{repsCross}  and \eqref{repsCrossS} we also see that 
$$ \pi_{\rho,x,s}(\sigma_{N,t}(f))= e^{it H}\, \pi_{\rho,x,s}(f) \, e^{-itH} \ \ \  \text{ and } \ \ \ 
\sigma_{N,t}(S_{N,k}) = e^{it H}\, S_{N,k} e^{-it H}\, .$$
We have
$$ Z_N(\beta)=\sum_{g\in \cS_{\rho,N}} |\det(g)|^{-\beta} \cdot \sum_{k=k_1\cdots k_n : k_i \in \Phi_N} k^{-\beta} $$
where $\Phi_N$ is the set of possible digits in the $N$-continued fraction expansion. 

\smallskip

For the first sum, we begin by considering the $N=1$ case. We now have that $\cS_{\rho}=\GL_2(\Z)\backslash M_2^\times(\Z)$,
where $M_2^\times(\Z)=\{ M \in M_2(\Z) \,|\, \det(M)\neq 0 \}$.
Thus, we are counting $\{ M\in M_2^\times(\Z)\,|\, |\det(M)|=n \}$ modulo $\GL_2(\Z)$. Up to a change
of basis in $\GL_2(\Z)$ we can always write a sublattice of $\Z^2$ in the form
$$ \begin{pmatrix} a & b \\ 0 & d \end{pmatrix} \Z^2 $$
with $a,d\geq 1$ and $0\leq b < d$, \cite{Serre}. 
Thus, we are equivalently counting such matrices with determinant $n$.
This counting is given by $\sigma(n)=\sum_{d|n} d$ so the first sum is
$\sum_{n\geq 1} \sigma(n)\, n^{-\beta} =\zeta(\beta)\zeta(\beta-1)$ as in the original $\GL_2$-system, and converges on $\beta \in (2, \infty )$. 

\smallskip

In the general case, we again consider $\rho\in \GL_2(\hat\Z)$, and by Lemma~\ref{SrhoNdet}
we now have that $\cS_{\rho,N}$ is the set of matrices in $M_2^\times(\Z)$ with determinant not divisible by any prime factor of $N$, up to the equivalence relation defined by $\GL_2(\Z)$. The first sum is then given by 
\begin{align*}
\sum_{g\in \cS_{\rho,N}} |\det(g)|^{-\beta} &= \sum_{n \geq 1  : (N,n)=1} \sigma(n) n^{-\beta} \\
&= \left( \sum_{n \geq 1  : (N,n)=1} n^{-\beta} \right) \left( \sum_{n \geq 1  : (N,n)=1} n^{-(\beta-1)}\right) \\
&= \zeta(\beta)\zeta(\beta-1) \prod_{p \; prime \; : \;p \mid N} \left(1 - p^{-\beta} \right) \left(1 - p^{-(\beta-1)}\right)
\end{align*}
where the counting $\sigma(n)=\sum_{d|n} d$ is identical to the $N=1$ case. Again, this series converges on $\beta \in (2, \infty)$.

To compute the second sum, let $P_{N,n}$ denote the total number of ordered factorizations of
$n$ into positive integer factors in $\alpha_N$. In the $N\geq1$ case, the sum we are considering is
\begin{align*}
 \sum_{n\geq 1} P_{N,n} \, n^{-\beta} &=  \sum_{k=1}^\infty \sum_{n=1}^\infty n^{-\beta} \sum_{n=n_1\cdots n_k : n_i \geq N} 1 & \\
 &=  \sum_{k\geq 1} \prod_{i=1}^k (\sum_{n_i\geq N} n_i^{-\beta}) & \\
  &= 
\begin{dcases}
\sum_{k=1}^\infty (\zeta(\beta))^k = \frac{1}{1-\zeta(\beta)}  &\textrm{if }N=1\\
\sum_{k=1}^\infty (\zeta(\beta) - \sum_{n=1}^{N-1} n^{-\beta})^k = \frac{1}{1 + \sum_{n=1}^{N-1} n^{-\beta} -\zeta(\beta) }  & \textrm{if } N>1. 
\end{dcases}
\end{align*}
In the $N=1$ case, note that the series $ \sum_{k=1}^\infty (\zeta(\beta))^k$ converges when $|\zeta(\beta)| < 1$. However, when $\beta >1$, $\zeta(\beta) > 1$ and the series does not converge there. The first series $\sum_{n\geq 1} \sigma_1(n)\, n^{-\beta} = \zeta(\beta)\zeta(\beta-1)$ converges for $\beta > 2$. Since the second series does not converge anywhere in the region $(2,\infty)$, there is no partition function. 

\smallskip

In the $N>1$ case, the relevant series converges when 
\begin{align*}
|\zeta(\beta) - \sum_{n=1}^{N-1} n^{-\beta}| = |\zeta(\beta) - (1 + \xi(\beta))| < 1
\end{align*}
 where $\xi(\beta) = 0$ when $N=2$ and $\xi(\beta) = \sum_{n=2}^{N-1} n^{-\beta}$ when $N>2$.  In 
the range $\beta \in (1,\infty)$ the $\zeta$-function is decreasing to $1$ and it crosses the value $\zeta(\beta)=2$
at a point  $\beta_{2,c} \sim 1.728647$. When $N=2$, the series converges on $(\beta_{2,c}, \infty)$. When $N>2$ we consider the  function $\zeta(\beta) - \xi(\beta)$ where $\xi(\beta)$ consists of a finite sum of terms of the form $n^{-\beta}$ each of which are continuous, decreasing to $0$ as $\beta \rightarrow \infty$ and have some finite value at $\beta=1$. Since $\lim_{\beta \rightarrow \infty} \zeta(\beta) - \xi(\beta) =1$ and $\lim_{\beta \rightarrow 1^+} \zeta(\beta) - \xi(\beta) = \infty$, there will be some point $\beta_{N,c}>1$ at which $\zeta(\beta_{N,c}) - \xi(\beta_{N,c})=2$. The corresponding series then converges on $(\beta_{N,c}, \infty)$. Since each $n^{-\beta}$ term is decreasing, we also know that $\beta_{N+1,c} < \beta_{N,c}$. 

\smallskip 

Similarly, in the $N\leq -1$ case we have
\begin{align*}
 \sum_{n\geq 1} P_{N,n} \, n^{-\beta} &=  \sum_{k=1}^\infty \sum_{n=1}^\infty n^{-\beta} \sum_{n=n_1\cdots n_k : n_i \geq |N|+1} 1  \\
 &=  \sum_{k\geq 1} \prod_{i=1}^k (\sum_{n_i\geq |N|+1} n_i^{-\beta})  \\
  &= \sum_{k=1}^\infty (\zeta(\beta) - \sum_{n=1}^{|N|} n^{-\beta})^k = \frac{1}{1 + \sum_{n=1}^{|N|} n^{-\beta} -\zeta(\beta) }  . 
\end{align*}
As before, this series converges on $(\beta_{N,c}, \infty)$ where for $N\leq-1$, $\beta_{N,c} = \beta_{1-N, c}$. In particular, $\beta_{-1,c}=\beta_{2,c}\sim 1.728647$. 
\endproof

Note that in the proof above we have shown that $\beta_{N,c}$ is decreasing in $N$ for positive $N$, and therefore attains its maximum value at $\beta_{2,c} = \beta_{-1, c}\sim 1.728647$. We also know that $\beta_{N,c}>1$ for all $N$. 

 \begin{figure}[H]
\centering
\begin{tikzpicture}
\draw[thick](-0.25,0)--(12.25,0);
\node() at (12,0){$|$};
\node[below]() at (12,-0.2){$2$};
\node() at (0,0){$|$};
\node[below]() at (0,-0.2){$1$};
\node[circle, draw, inner sep=1.5pt, fill=black]() at (8.74,0){};
\node[below]() at (8.74,0){\small{$\beta_{2,c}$}};
\node[circle, draw, inner sep=1.5pt, fill=black]() at (6.993469166,0){};
\node[below]() at (6.993469166,0){\small{$\beta_{3,c}$}};
\node[circle, draw, inner sep=1.5pt, fill=black]() at (6.23944128,0){};
\node[below]() at (6.23944128,0){\small{$\beta_{4,c}$}};
\node[circle, draw, inner sep=1pt, fill=black]() at (5.795370051,0){};
\node[circle, draw, inner sep=1pt, fill=black]() at (5.493528774,0){};
\node[circle, draw, inner sep=1pt, fill=black]() at (5.270645375,0){};
\node[circle, draw, inner sep=1pt, fill=black]() at (5.0969396,0){};
\node[circle, draw, inner sep=1pt, fill=black]() at (4.956328433,0){};
\node[circle, draw, inner sep=1.5pt, fill=black]() at (4.839265134,0){};
\node[below]() at (4.839265134,0){\small{$\beta_{10,c}$}};
\node[circle, draw, inner sep=1.5pt, fill=black]() at (3.33792,0){};
\node[below]() at (3.33792,0){\small{$\beta_{10^2,c}$}};
\node[circle, draw, inner sep=1pt, fill=black]() at (2.63424,0){};
\node[circle, draw, inner sep=1pt, fill=black]() at (2.20644,0){};
\node[circle, draw, inner sep=1pt, fill=black]() at (1.91364,0){};
\node[circle, draw, inner sep=1pt, fill=black]() at (1.69836,0){};
\node[circle, draw, inner sep=1pt, fill=black]() at (1.53228,0){};
\node[circle, draw, inner sep=1pt, fill=black]() at (1.39968,0){};
\node[circle, draw, inner sep=1pt, fill=black]() at (1.29096,0){};
\node[circle, draw, inner sep=1.5pt, fill=black]() at (1.2,0){};
\node[below]() at (1.2,0){\small{$\beta_{10^{10},c}$}};
\end{tikzpicture}
\caption{$N$-dependence of Boundary-$\GL_2$ critical temperature ($\beta_{N,c}$)}
\end{figure}

\begin{lem}\label{ZbetaPoles}
For $N \geq 2$ and $N \leq -1$, the partition function $Z_N(\beta)$ of Proposition~\ref{timeevolZ} is defined by an absolutely
convergent series 
\begin{align*}Z_N(\beta)=\Tr(e^{-\beta H_N})=\sum_{\lambda\in \Sp(H_N)} e^{-\beta \lambda}
\end{align*}
for $\beta >2$. Its analytic continuation \eqref{Zetabeta} has poles at $\beta\in \{ 1, \beta_{N,c} , 2 \}$,
for a point $1< \beta_{N,c}<2$. In the geometrically relevant case of $N=-1$, $\beta_{-1,c} \sim 1.728647$.
\end{lem}

\proof As argued in the proof of \ref{timeevolZ}, the denominator of $Z_N(\beta)$ has a single zero at a point $1<\beta_{N,c}<2$. The Riemann zeta function $\zeta(\beta)$ has a pole at $\beta=1$. Therefore, the sum $\sum_{n\geq 1 \; : \; (N,n)=1} \sigma_1(n) n^{-\beta}$ is
convergent for $\beta>2$ and its analytic continuation $\zeta(\beta)\zeta(\beta-1) \prod_{p \; prime \; : \;p \mid N} \left(1 - p^{-\beta} \right) \left(1 - p^{-(\beta-1)}\right)$ has
poles at $\beta=2$ and $\beta=1$. 
\endproof  

\smallskip
\subsection{KMS states.}

We classify the KMS states for the family of dynamical systems $(\cA_{\partial, G,\cP_N}, \sigma_{N,t})$. Since we have $\cA_{\partial, G,\cP_N} =\cB_{\partial,N} \rtimes {\rm Red_N}$, we consider separately the KMS states for the modified $\GL_2$ part $\cB_{\partial,N}$ of the system, and the part of the system generated by the isometries $S_{N,k}$ in the semigroup $\textrm{Red}_N$, which is a Cuntz-Krieger-Toeplitz type algebra. The KMS states of the Cuntz-Krieger-Toeplitz type algebras have been studied by \cite{ExelLaca}, and we draw on their main results. We show that in the $N=1$ case, corresponding the the standard $\GL_2$ system, there are no KMS states at any temperature, though we may still define a ground state. In all other cases, the system has two critical temperatures at $\beta = \beta_{N,c}<2$ and $\beta=2$. When $\beta < \beta_{N,c}$ there are no KMS$_{\beta}$ states. When $\beta_{N,c}<\beta<2$, the structure of the KMS$_{\beta}$ states will be identical to the structure on the modified $\GL_2$ part of the system alone, though we have not computed this explicitly. When $\beta >2$, the KMS$_{\beta}$ states are given by Gibbs states, whose limit as $\beta \rightarrow \infty$ gives the ground states. 

\begin{lem}\label{CKTalgebra}
The subalgebra of $\cA_{\partial, G,\cP_N}$ generated by the family $\{S_{N,k}\}_{k \in \alpha_N}$ of isometries  is a Cuntz-Krieger-Toeplitz algebra, denoted by $\mathcal{O}_A$ in the setting of \cite{ExelLaca} with the infinite matrix $A=\{A(x,y)\}_{x,y \in \alpha_N}$ given by $A(x,y)=1$ for all $x,y \in \Phi_N$. In other words, it satisfies the following properties on its initial and final projections $q_{k} = S_{N,k}^*S_{N,k}$ and $p_{k}=S_{N,k}S_{N,k}^*$. For all $k,j \in \Phi_N$: 
\begin{enumerate}
\item $q_{k}q_{j} = q_{j}q_{k}$, 
\item $S_{N,k}^*S_{N,j} =0 \; \textrm{if} \; j \neq k$, 
\item $q_k S_{N,j} = S_{N,j}$, 
\item and $\prod_{k \in X} q_k \prod_{j \in Y} (1 - q_j) = 0$ for $X,Y$ finite subsets of $\Phi_N$. 
\end{enumerate}
\end{lem}

\begin{proof}
Conditions (1), (3), and (4) follow from the fact that $S_{N,k}^*S_{N,k} =1$ (Lemma \ref{Tactiongk}). Condition (2) is easily verified. If $k \neq j$ then 
\begin{align*}
S_{N,k}^*S_{N,j} \xi(x) &= S_{N,k}^* ( \chi_{X_{N,j}}(x) \xi(g_{N,j}^{-1} x) ) \\
&= \chi_{X_{N,j}}(g_{N,k}x) \xi (g_{N,k}g_{N,j}^{-1} x) = 0. 
\end{align*}
\end{proof}

\begin{prop}\label{KMSstates}
The KMS$_{\beta}$ states of the dynamical system $(\cA_{\partial, G,\cP_N}, \sigma_{N,t})$ can be characterized as follows. When $N=1$, there are no KMS$_{\beta}$ states for any temperature $\beta$.  When $N\leq -1$ or $N>1$, the system has a critical temperature $\beta_{N,c} \in (1,2)$. We then have the following. 
\begin{enumerate}
\item When $\beta < \beta_{N,c}$ there are no $\textrm{KMS}_{\beta}$ states. 
\item When $\beta > \beta_{N,c}$, there is one KMS$_{\beta}$ state for every KMS$_{\beta}$ state of the modified $\GL_2$ system corresponding to $\cB_{\partial, N}$. 
\item When $\beta > 2$, the KMS$_{\beta}$ states restrict to the $\textrm{Red}_N$ part of the system
as the unique KMS$_{\beta}$ state of the Cuntz-Krieger-Toeplitz algebra and restrict to a KMS$_{\beta}$
state on the $\cB_{\partial,N}$ part of the system. All the extremal KMS$_{\beta}$ states are Gibbs. In particular, one obtains a family of extremal KMS-states
corresponding to the Gibbs states of $\cB_{\partial,N}$, 
parameterized by the set
$$ \cE_\beta \cong \left\{ \begin{array}{ll} \displaystyle{ \GL_2(\Z) \backslash (\GL_2(\hat{\Z}) \times \tilde\cP_N ) \times \cD_{[0,1]\cap \Q} } & N \neq -1 \\ 
\displaystyle{ \SL_2(\Z) \backslash (\GL_2(\hat{\Z}) \times \tilde\cP_{-1} ) \times \cD_{[0,1]\cap \Q} } & N = -1 \, . \end{array} \right. $$
Note that although there are no KMS$_{\beta}$ states when $N=1$, in this case $\cE_{\beta}$ still parameterizes a family of ground states at $\beta=\infty$, see Remark \ref{ground}. 
\end{enumerate}
\end{prop}

\proof
If there is a KMS$_{\beta}$ state on$(\cA_{\partial, G,\cP_N}, \sigma_{N,t})$, it must restrict to a KMS$_{\beta}$ state on the subalgebra of $\cA_{\partial, G,\cP_N}$ generated by the family of isometries $\{S_{N,k}\}_{k \in \alpha_N}$. We will first characterize the KMS$_{\beta}$ states of this subalgebra, which we have established in Lemma \ref{CKTalgebra} is a Cuntz-Krieger-Toeplitz algebra. We also note that since $\Phi_N$ is a countable set and in our case the matrix $A$ is simply a matrix with every entry set to $1$, Standing Hypothesis 8.1 (i), and (ii) of \cite{ExelLaca} are satisfied. The dynamics $\sigma_{N,t}$ on the subalgebra take the form $\sigma_{N,t}(S_{N,k})= k^{it} S_{N,k}$ for each $k \in \Phi_N$, and since $\Phi_N$ is a set of real numbers in the interval $(1,\infty)$ the rest of Standing Hypothesis 8.1 of \cite{ExelLaca} is also satisfied. 

\smallskip

We now draw on the main results of \cite{ExelLaca}. Corollary 9.7 states that there is a critical temperature $\dot \beta_{N,c}$ above which there is a single KMS$_{\beta}$ state at each temperature $\beta$. Theorem 14.5 states that there is a second critical temperature $\ddot \beta_{N,c}$ below which there are no KMS$_{\beta}$ states at all. These critical temperatures are defined as follows. Let $\Omega_N$ be the set of words in $\Phi_N$ and $\Omega_{N,j,k}$ be the set of words in $\Phi_N$ beginning with $j$ and ending with $k$. Then $\dot \beta_{N,c}$  and $\ddot \beta_{N,c}$  are the abscissas of convergence of the series 
\[ Z(\beta) = \sum_{\mu \in \Omega_N} (\mu)^{-\beta} \;\;\; \textrm{and} \;\;\; Z_{jk}(\beta) = \sum_{\mu \in \Omega_{N,j,k}} (\mu)^{-\beta}\] 
respectively. Note that the abscissa of convergence of the second series is independent of the choice of $j$ and $k$. We will now calculate these critical temperatures. 

\smallskip

The partition function $Z(\beta)$ has already been calculated in the second part of Proposition \ref{timeevolZ} and is given by

\begin{align*}
 Z(\beta) = \begin{dcases}
\sum_{k=1}^\infty (\zeta(\beta))^k  &\textrm{if }N=1\\
\sum_{k=1}^\infty (\zeta(\beta) - \sum_{n=1}^{N-1} n^{-\beta})^k  & \textrm{if } N>1 \\
 \sum_{k=1}^\infty (\zeta(\beta) - \sum_{n=1}^{|N|} n^{-\beta})^k& \textrm{if } N \leq -1. \\
\end{dcases}
\end{align*}

Modifying this calculation slightly we find that 
\begin{align*}
Z_{jk} (\beta) &= (jk)^{-\beta} + \sum_{n=1}^{\infty} \sum_{\mu \in \Omega_N : |\mu| = n} j^{-\beta} \mu^{-\beta} k^{-\beta} \\
& = (jk)^{-\beta} \left( 1 +  \sum_{n=1}^{\infty} \prod_{i=1}^n \sum_{k_i \in \alpha_N} k_i^{-\beta}\right) = (jk)^{-\beta} Z(\beta).
\end{align*}

Clearly $Z(\beta)$ and $Z_{jk}(\beta)$ have the same abscissa of convergence, $\ddot \beta_{N,c} = \dot\beta_{N,c}$. In fact in the $N \neq 1$ case, this abscissa of convergence is $\beta_{N,c}$ of \ref{timeevolZ}. When $N=1$, neither series converges for any value of $\beta$, $\dot \beta_{N,c} = \ddot \beta_{N,c} = \infty$. Hence there are no KMS$_{\beta}$ states for any finite inverse temperature $\beta$. 
\smallskip

Now we focus our attention on the subalgebra corresponding to $\cB_{\partial, N}$ in the $N \neq 1$ case. In the range $\beta >2$, $e^{-\beta H_N}$ is trace class, by Lemma \ref{ZbetaPoles}. We therefore have Gibbs states of the form, for $X\in \cA_{\partial,G,\cP_N}$

$$
 \varphi_{\beta,N}( X )  =\frac{\Tr( \pi_{\rho,x,s}(X) e^{-\beta H_N })}{\Tr(e^{-\beta H_N}) } $$
$$ = Z_N(\beta)^{-1} \sum_{k_i \in \alpha_N, g\in \cS_{\rho,N}} |\det(g)|^{-\beta} (k_1\cdots k_n)^{-\beta}\,  \langle \delta_g \otimes \epsilon_{k_1,\ldots, k_n}, \pi_{\rho,x,s} (X) \delta_g \otimes \epsilon_{k_1,\ldots, k_n}\rangle. $$
 depending on our choice of representation $\pi_{\rho,x,s}$. 
 \smallskip
 
We now restrict to the subalgebra $\cB_{\partial,N}$. The Gibbs states above give, for $f \in \cB_{\partial,N}$
 \begin{align*}
 \varphi_{\beta,\rho,x,s}(f) & = \frac{\Tr( \pi_{\rho,x,s}(f) e^{-\beta H_N })}{\Tr(e^{-\beta H_N}) } \\
 &= Z(\beta)^{-1} \sum_{k_i \in \alpha_N, g\in\cS_{\rho,N}} |\det(g)|^{-\beta} (k_1\cdots k_n)^{-\beta} f(1,g\rho, g_\gamma(x,gs))
 \end{align*} 
where $\gamma \in \textrm{Red}_N$ is determined by $k_1,...,k_n$. 
Thus, the Gibbs states are parameterized by the choice of $\rho \in \GL_2(\hat{\Z})$ and $s\in \tilde\cP_N$, up
to the $\Gamma_N$ equivalence, and by the choice of the point $x\in \cD_{[0,1]\cap \Q}$.
By Lemma~\ref{GammaNinvrho} we know
that two $\rho,\rho'\in \GL_2(\Z)$ are in the same orbit of the $\Gamma_N$-action,
$\rho'=\gamma\rho$, with $\gamma\in \Gamma_N$ iff $\gamma\in \GL_2(\Z)$.
Thus, the parameterizing space of the low-temperature extremal KMS$_\beta$ states
is given by the quotient 
\begin{equation}\label{EbetaN}
\cE_\beta \cong \GL_2(\Z)\backslash (\GL_2(\hat\Z)\times \tilde\cP_N) \times \cD_{[0,1]\cap \Q} \, , 
\end{equation}
for all $N\neq 1$ and by $\SL_2(\Z)\backslash (\GL_2(\hat\Z)\times \tilde\cP_{-1}) 
\times \cD_{[0,1]\cap \Q}$ in the case $N=-1$.

Finally, we note that when $\beta >2$, all the extremal KMS$_{\beta}$ states are Gibbs as a consequence of the convergence of the partition function. An extremal KMS state is factor (see e.g. \cite{BR2} Theorem 5.3.30 ) and thus can be written as trace against a density matrix in the enveloping von Neumann algebra (see e.g. \cite{BR1} Theorem 2.4.1.) When $\beta >2$, $e^{-\beta H_N}$ is traceclass by Lemma \ref{ZbetaPoles}, and will, after normalization, be the required density matrix. 
\endproof

Figure~\ref{FigKMS} compares the classification of the KMS states of the standard Bost-Connes system and of the $\GL_2$ system with the newly constructed boundary-$\GL_2$-systems. 

 \begin{figure}[H] \label{figkms}
\centering
\begin{tikzpicture}
\begin{scope}[scale=0.9]
\node[]() at (0,1.5){Bost-Connes and $\GL_2$-systems};
\node[left]() at (-6.2,0){BC};
\node[left]() at (-6.2,-2){$\GL_2$};
\draw[thick, -stealth] (-6,0)--(7,0);
\node[below]() at (-6,-0.25){$0$};
\node[below]() at (-1.5,-0.25){$1$};
\node[above]() at (-3.75,0){\small{unique $\beta$-KMS states}};
\node[above]() at (3,0){\small extremal $\beta$-KMS param. by {$ \GL_1(\hat\Z)$}};
\node() at (-1.5,0){$]$};
\node() at (-6,0){$]$};
\begin{scope}[shift={(0,-2)}]
\draw[thick,-stealth] (-6,0)--(7,0);
\node[below]() at (-6,-0.25){$0$};
\node[below]() at (-1.5,-0.25){$1$};
\node[below]() at (3,-0.25){$2$};
\node[above]() at (-3.75,0){\small{no $\beta$-KMS states}};
\node[]() at (-3.75,0){$\times$};
\node[above]() at (0.75,0){\small{unique $\beta$-KMS states}};
\node[above]() at (5,0.55){\qquad\small extremal $\beta$-KMS param. };
\node[above]() at (5,0){\qquad\small by {$\SL_2(\bZ) \backslash (\GL_2(\hat\Z)\times \H)$}};
\node() at (-1.5,0){$]$};
\node() at (3,0){$]$};
\node() at (-6,0){$]$};
\end{scope}
\begin{scope}[shift={(0,-5)}]
\node[left]() at (-6.2,0){$N=1$};
\node[]() at (0,1.5){Boundary-$\GL_2$-systems};
\draw[thick,-stealth] (-6,0)--(7,0);
\node[below]() at (-6,-0.25){$0$};
\node[above]() at (0.5,0){\small{no $\beta$-KMS states}};
\node[]() at (0.5,0){$\times$};
\node() at (-6,0){$]$};
\begin{scope}[shift={(0,-2.4)}]
\draw[thick,-stealth] (-6,0)--(7,0);
\node[below]() at (-6,-0.25){$0$};
\node[below]() at (1.5,-0.25){\color{gray}{$\beta_{-1,c}$}};
\node[below]() at (1.5,-0.8){\tiny{\color{gray}{$(\simeq 1.728647)$}}};
\node[below]() at (3,-0.25){$2$};
\node[above]() at (-3,0){\small{no $\beta$-KMS states}};
\node[]() at (-3,0){$\times$};
\node[above]() at (4.8,1.1){\qquad\small subset of extremal $\beta$-KMS param. by };
\node[above]() at (4.8,0.5){\qquad\small  {$\Gamma_{\pm1}\backslash (\GL_2(\hat\Z)\times \tilde\cP_N) \times \cD_{[0,1]\cap \Q}$}};
\node[below]() at (-0.5,-0.25){$\beta_{N,c}$};
\node[below]() at (-0.5,-0.8){\tiny{$\in (1,\beta_{-1,c}]$}};
\node() at (1.8,0){\color{gray}{$]$}};
\node() at (-0.5,0){$]$};
\node() at (3,0){$]$};
\node() at (-6,0){$]$};
\draw [decorate,
    decoration = {calligraphic brace, mirror, amplitude=6pt}, thick] (-0.5,-1.4) --  (7,-1.4);
    \draw [decorate,
    decoration = {calligraphic brace, amplitude=6pt}, thick] (3,0.4) --  (7,0.4);
\node[below]() at (3.5,-1.5){\small{$\beta$-KMS states mirror modified $\GL_2$-system}};
\node[left]() at (-6.2,0){$N \neq 1$};
\draw[- stealth](0.8,-0.6)--(0,-0.6);
\end{scope}
\end{scope}
\end{scope}
\end{tikzpicture}
\caption{KMS States of the Bost-Connes, $\GL_2$, and Boundary-$\GL_2$-systems. \label{FigKMS}}
\end{figure}

The precise structure of the KMS states in the region $\beta \in (\beta_{N,c},2)$ remains an open problem. The standard $\GL_2$ system (when $\Gamma = \SL_2(\Z)$) has been studied in \cite{LLN2}. Their analysis of the behavior when $\beta \in (1,2)$ is not directly applicable in our case, because for $N=1$, the Red$_N$ part of the system has no KMS states at any temperature $\beta$. However, it would be interesting to see whether a similar analysis can be applied when $\Gamma = \Gamma_N$ for some $N>1$ or $N \leq -1$.

\smallskip

Furthermore, for $\beta >2$, we focused on exhibiting a family of extremal KMS states that directly
generalize those of the $\GL_2$-system, but we have not shown that the Gibbs states of the form $ \varphi_{\beta,\rho,x,s}$ parameterize the entire set of extremal points in the simplex of KMS$_{\beta}$ states, only that all such states are Gibbs and that these ones in particular are defined. A full characterization of the extremal KMS states along the lines of what was done for the $\GL_2$ system in Theorem 1.26 of \cite{CoMa2} may be possible, but we leave it to a future work.

\smallskip

As in \cite{CoMa} we consider the ground states at zero temperature as the
weak limit of the Gibbs states for $\beta\to \infty$
$$ \varphi_{\infty,\rho,x,s}(f) =\lim_{\beta \to \infty} \varphi_{\beta,\rho,x,s}(f). $$
\begin{cor}
When $N\neq1$, the ground states are given by 
$$ \varphi_{\infty,\rho,x,s}(f) = f(1,\rho,x,s).$$
\end{cor}
\proof
Observe that whenever $N\neq1$, $\lim_{\beta \rightarrow \infty} Z_N(\beta) = 1$. Furthermore, the only $g \in \cS_{\rho,N}$ with $|\det(g)|=1$ is the identity element, and hence the only terms for which $\lim_{\beta \rightarrow \infty} |\det(g)|^{-\beta}$ does not vanish are those for which $g = 1$. A word in $\alpha_N$ satisfies $k_1... k_n \geq |N|^n$  if $N>1$ and $k_1 ... k_n \geq (|N| +1)^n$ if $N \leq 1$. Hence $\lim_{\beta \rightarrow \infty} |k_1...k_n|^{-\beta}$ vanishes unless $k_1...k_n$ is the empty word. We have that the ground states are 
\begin{align}
\label{groundstate}
\varphi_{\infty,\rho,x,s}(f) = \langle \delta_1 \otimes \epsilon_\emptyset, \pi_{\rho,x,s}(f)\, \delta_1 \otimes \epsilon_\emptyset\rangle =f(1,\rho,x,s),
\end{align}
the evaluation of the function $f$ at the point $g=1$ and $(\rho,x,s)$ that determines the representation
$\pi_{\rho,x,s}$. 
\endproof

\begin{rem}\label{ground}
Although in the $N=1$ case there are no low-temperature KMS states, and hence the weak limit does not exist, we can still define the projection onto the kernel of the Hamiltonian as in \ref{groundstate}. This satisfies the weak KMS condition in the sense that function 
\[ F(t) = \varphi_{\infty, \rho, x,s} (f \sigma_t(f')) = f(1,\rho,x,s) f'(1, \rho, x,s)\]
has a bounded holomorphic extension to the upper half plane.
\end{rem} 

\medskip
\section{Averaging on geodesics and boundary values}

In this section we construct boundary values of the observables of the
$\GL_2$-system of \S \ref{GL2Sec}. We use the theory of limiting modular
symbols of  \cite{ManMar}. We show that the resulting boundary values localize
nontrivially at the quadratic irrationalities and at the level sets of the
multifractal decomposition considered in \cite{KesStra}. We discuss 
in particular the case of quadratic irrationalities.

\subsection{Geodesics between cusps}

Let $C_{\alpha,\epsilon,s}$ with $\alpha \in \P^1(\Q)\smallsetminus \{ 0 \}$, $\epsilon\in \{ \pm \}$,
and $s\in \P$ denote the geodesic in $\H^\epsilon \times \P$ with endpoints at the cusps 
$(0,s)$ and $(\alpha,s)$ in $\P^1(\Q)\times \P$. 

\smallskip

Let $p_k(\alpha), q_k(\alpha)$ be the successive numerators and denominators of the 
$\GL_2(\Z)$-continued fraction expansion of $\alpha\in \Q$ with $p_n(\alpha)/q_n(\alpha)=\alpha$ and let
$$ g_k(\alpha):=\begin{pmatrix} p_{k-1}(\alpha) & p_k(\alpha) \\ q_{k-1}(\alpha) & q_k(\alpha) \end{pmatrix} \in \GL_2(\Z). $$ 

\smallskip

We denote by $C^k_{\alpha,\epsilon,s}$ the geodesic in $\H^\epsilon \times \P$ with endpoints
at the cusps 
$$ \frac{p_{k-1}(\alpha)}{q_{k-1}(\alpha)} = g_k^{-1}(\alpha)\cdot 0 \ \ \text{ and } \ \ 
\frac{p_k(\alpha)}{q_k(\alpha)} = g_k^{-1}(\alpha)\cdot \infty , $$
where $g\cdot z$ for $g\in \GL_2(\Z)$ and $z\in \H^\pm$ is the action by fractional linear transformations.

\smallskip

For $C$ a geodesic in $\H^\pm$ let $ds_C$ denote the geodesic length element. In the case of
$C_{\infty,\epsilon,s}$ we have $ds_{C_{\infty,\epsilon,s}}(z)=dz/z$. 

\smallskip

We use the notation $\{ g\cdot 0, g\cdot \alpha \}_G$ to denote the homology class determined by
the image in the quotient $X_G=G\backslash \H = \GL_2(\Z)\backslash (\H^\pm \times \P)$ 
of the geodesic $C_{\alpha,\epsilon,s}$, for $g$ a representative of $s\in \P$. Similarly we
write $\{ \alpha, \beta \}_G$ for homology classes determined by the images in the
quotient $X_G$ of geodesics in $\H^\pm \times \P$ with endpoints $\alpha,\beta$ at cusps
in $\P^1(\Q)\times \P$.

\smallskip
\subsection{Limiting Modular Symbols}

We recall briefly the construction of limiting modular symbols from \cite{ManMar}.
We consider here some finite index subgroup $G\subset \Gamma$
of $\Gamma=\PGL_2(\Z)$. We denote by $\P= \Gamma/G$ the finite coset space of this subgroup. 
We also write the quotient modular curve as $X_G=G\backslash \H = \Gamma\backslash (\H \times \P)$.

\smallskip

Recall that the classical modular symbols $\{ \alpha, \beta \}_G$, with $\alpha,\beta\in \P^1(\Q)$
are defined as the homology classes in $H_1(X_G,\R)$ defined as functionals that integrate
lifts to $\H$ of differentials on $X_G$ along the geodesic arc in $\H$ connecting $\alpha$ and $\beta$, see
\cite{Man-modsymb}.
They satisfy additivity and invariance: for all $\alpha,\beta,\gamma \in  \P^1(\Q)$ 
$$ \{ \alpha,\beta \}_G + \{ \beta,\gamma \}_G =\{ \alpha, \gamma\}_G \ \ \text{ and } \ \ 
 \{ g\alpha, g\beta \}_G =\{ \alpha, \beta \}_G \ \  \forall g\in G. $$
Thus, it suffices to consider the modular symbols of the form $\{ 0, \alpha \}_G$ with $\alpha\in \Q$. 
These satisfy the relation
 $$ \{ 0, \alpha \}_G = - \sum_{k=1}^n  \{ \frac{p_{k-1}(\alpha)}{q_{k-1}(\alpha)},  \frac{p_k(\alpha)}{q_k(\alpha)} \}_G =
- \sum_{k=1}^n  \{ g_k^{-1}(\alpha)\cdot 0, g_k^{-1}(\alpha)\cdot \infty \}_G , $$
 where $p_k(\alpha), q_k(\alpha)$ are the successive numerators and denominators of the 
 $\GL_2(\Z)$-continued fraction expansion of $\alpha\in \Q$ with $p_n(\alpha)/q_n(\alpha)=\alpha$ and
 $$ g_k(\alpha)=\begin{pmatrix} p_{k-1}(\alpha) & p_k(\alpha) \\ q_{k-1}(\alpha) & q_k(\alpha) \end{pmatrix} . $$ 
 (There is an analogous formula for the $\SL_2(\Z)$-continued fraction.)
 
 \smallskip
 
 Limiting modular symbols were introduced in \cite{ManMar}, to account for the noncommutative
 compactification of the modular curves $X_G$ by the boundary $\P^1(\R)$ with the $G$ action. 
 One considers the infinite geodesics $L_\theta=\{ \infty, \theta \}$ given by the vertical lines
 $L_\theta =\{ z\in \H \,|\, \Re(z)=\theta \}$ oriented from the point at infinity to the point $\theta$ on
 the real line. Upon choosing an arbitrary base point $x\in L_\beta$ let $x(s)$ denote the point on
 $L_\beta$ at an arc-length distance $s$ from $x$ in the orientation direction. One considers the
 homology class $\{ x, x(s) \}_G \in H_1(X_G,\R)$ determined by the geodesic arc between $x$ and $x(s)$.
 The limiting modular symbol is defined as the limit (when it exists) 
 \begin{equation}\label{limmodsymb}
 \{\{ \star, \theta \}\}_G := \lim_{s\to \infty} \frac{1}{s} \{ x, x(s) \}_G \in H_1(X_G,\R). 
 \end{equation}
 
 \smallskip
 
 It was shown in \cite{ManMar} that the limit \eqref{limmodsymb} exists on a full measure set 
 and can be computed by an ergodic average. More generally, it was shown in \cite{Mar} that
 there is a multifractal decomposition of the real line by level sets of the Lyapunov exponent
 of the shift of the continued fraction expansion plus an exceptional set where the limit does
 not exist. On the level sets of the Lyapunov exponent the limit is again given by an average
 of modular symbols associated to the successive terms of the continued fraction expansion.
 More precisely, as in the previous section, let $T:[0,1] \to [0,1]$ denote the shift map of the 
 continued fraction expansion,
 $$ T(x)=\frac{1}{x} - \left[ \frac{1}{x} \right], $$
 extended to a map $T: [0,1]\times \P \to [0,1]\times \P$ with $\P=\Gamma/G$.
 The Lyapunov exponent of the shift map is given by the limit (when it exists)
 \begin{equation}\label{Lyapexp}
  \lambda(x) = \lim_{n\to \infty} \frac{1}{n} \log |(T^n)^\prime(x)| = 2\lim_{n\to \infty} \frac{1}{n} \log q_n(x) , 
 \end{equation} 
 where $q_n(x)$ are the successive denominators of the continued fraction expansion of $x\in [0,1]$. There
 is a decomposition $[0,1]=\cup_\lambda \cL_\lambda \cup \cL'$ where 
 $\cL_\lambda=\{ x\in [0,1]\,|\, \lambda(x)=\lambda \}$ and $\cL'$ the set on which the limit \eqref{Lyapexp}
 does not exist. For all $\theta\in \cL_\lambda$ the limiting modular symbol \eqref{limmodsymb} is then given by
 \begin{equation}\label{lms2}
 \{\{ \star, \theta \}\}_G =\lim_{n\to \infty} \frac{1}{\lambda n} \sum_{k=1}^n \{ g_k^{-1}(\theta)\cdot 0, g_k^{-1}(\theta) \cdot \infty \}_G = \lim_{n\to \infty} \frac{1}{\lambda n} \sum_{k=1}^n \{ \frac{p_{k-1}(\theta)}{q_{k-1}(\theta)}, \frac{p_k(\theta)}{q_k(\theta)} \}_G .
 \end{equation}
 
 \smallskip
 
 The results of \cite{ManMar} and \cite{Mar} show that the limiting modular symbol \eqref{lms2}
 vanishes almost everywhere, with respect to the Hausdorff measure of $\cL_\lambda$. 
 However, non-vanishing values of the limiting modular symbols are obtained, for example, 
 for all the quadratic irrationalities. 
 
 \smallskip
 
 In the case of quadratic irrationalities, one obtains two equivalent descriptions of the limiting
 modular symbol, one that corresponds to integration on the closed geodesic $C_\theta =\Gamma_\theta\backslash S_\theta$ with $S_\theta$ the infinite geodesic with endpoints the Galois conjugate pair $\theta,\theta'$ and
 the other in terms of averaged integration on the modular symbols associated to the (eventually periodic) 
 continued fraction expansion. We obtain the identification of homology classes in $H_1(X_G,\R)$
 \begin{equation}\label{lmsRM}
  \{\{ \star, \theta \}\}_G = \frac{\sum_{k=1}^\ell \{ \frac{p_{k-1}(\theta)}{q_{k-1}(\theta)}, \frac{p_k(\theta)}{q_k(\theta)} \}_G}{\lambda(\theta) \ell } = \frac{\{ 0, g\cdot 0 \}_G}{\ell(g)}  \, \in H_1(X_G,\R).
 \end{equation}
 In the first expression $\ell$ is the minimal positive integer for which $T^\ell (\theta) =\theta$
 and the limit defining the Lyapunov exponent $\lambda(\theta)$ exists for quadratic irrationalities. 
 In the second expression $g\in \Gamma$ is the hyperbolic generator of $\Gamma_\theta$ with
 fixed points $\theta,\theta'$, with eigenvalues $\Lambda_g^\pm$ and 
 $\{ 0, g\cdot 0 \}_G$ denotes the homology class in $H_1(X_G,\R)$ of the closed geodesic $C_\theta$
 and $\ell(g)=\log \Lambda_g^-=2\log \epsilon$ is the length of $C_\theta$. 
 
 \smallskip
 
 A more complete analysis of the values of the limiting modular symbols was then carried out in
 \cite{KesStra}, where it was shown that, in fact, the limiting modular symbols is non-vanishing
 on a multifractal stratification of Cantor sets of positive Hausdorff dimension. We will recall more
 precisely this result in \S \ref{BoundValSec} below. 
 
 \smallskip
 
 The construction recalled above of limiting modular symbols determine non-trivial real homology
classes in the quotient $X_G$ associated to geodesics in $\H$ with endpoints in one of the
multifractal level sets of \cite{KesStra}.  
These homology classes pair with 1-forms on $X_G$, and
in particular with weight $2$ cusp forms for the finite index subgroup $G$.

\smallskip

Let $\cM_{G,k}$ the $\C$-vector space of modular forms of weight $k$ for the finite index subgroup $G\subset
\GL_2(\Z)$ and let $\cS_{G,k}$ be the subspace of cusp forms. Let $X_G=\GL_2(\Z)\backslash (\H^\pm \times \P)$
be the associated modular curve. We denote by 
\begin{equation}\label{pairSMod}
\langle\cdot, \cdot \rangle: \cS_{G,2} \times H_1(X_G,\R) \to \C
\end{equation}
the perfect pairing between cusp forms of weight two and modular symbols., which we equivalently write as 
integration
\begin{equation}\label{pairSMod2}
 \langle \psi, \{ \alpha, \beta \}_G \rangle = \int_{\{\alpha, \beta\}_G} \psi(z)\, dz. 
\end{equation} 

\medskip
\subsection{Boundary values}\label{BoundValSec}

We consider now a linear map, constructed using cusp forms and limiting modular symbols,
that assigns to an observable of the bulk $\GL_2$-system a boundary value.

\smallskip

Let $\cL\subset [0,1]$ denote the subset of points such that the Lyapunov exponent \eqref{Lyapexp}
of the shift of the continued fraction exists. The set $\cL$ is stratified by level sets $\cL_\lambda=\{
x\in [0,1]\,|\, \lambda(x)=\lambda \}$, with the Lyapunov spectrum given by the Hausdorff dimension 
function $\delta(\lambda)=\dim_H (\cL_\lambda)$. Recall also that, for a continuous function $\phi$
on a $T$-invariant subset $E \subset [0,1]$, the Birkhoff spectrum (see \cite{FanFeng}) is the function
\begin{equation}\label{BirSpec}
 f_E(\alpha):= \dim_H \cL_{\phi,E,\alpha},
\end{equation} 
where $\cL_{\phi,E,\alpha}$ are the level sets of the Birkhoff average by 
\begin{equation}\label{levelBirkhoff}
 \cL_{\phi,E,\alpha}:= \{ x \in E \,|\, \lim_{n\to \infty} \frac{1}{n}\sum_{k=0}^{n-1} \phi \circ T^k(x) =\alpha \}. 
\end{equation}
In particular, $\cL_\lambda=\cL_{\phi,\lambda}$ for $\phi(x)=\log |T^\prime(x)|$.
Lyapunov and Birkhoff spectra for the shift of the continued fraction expansion
are analyzed in \cite{PoWe}, \cite{FLWW}.

\smallskip

In a similar way, one can consider the multifractal spectrum associated to the level
sets of the limiting modular symbol, as analyzed in \cite{KesStra}. Let $f_1,\ldots, f_g$
be a basis of the complex vectors space $\cS_{G,2}$ of cusp forms of weight $2$ for
the finite index subgroup $G\subset \GL_2(\Z)$. Let $\Re(f_i)$, $\Im(f_i)$ be the
corresponding basis as a real $2g$-dimensional vector space. Under the pairing
\eqref{pairSMod}, \eqref{pairSMod2}, which identifies $\cS_{G,2}$ with the dual of
$H_1(X_G,\R)$, we can define as in \cite{KesStra} the level sets $E_\alpha$, 
for $\alpha\in \R^{2g}$, of the limiting modular symbol as 
\begin{equation}\label{EalphaLimMod}
E_\alpha:= \{ (x,s)\in [0,1]\times \P \,|\, ( \langle f_i, \{\{ \star, x \}\}_G \rangle =\alpha \in \R^{2g} \}. 
\end{equation}
Equivalently, we write this as
$$ E_\alpha =\{ (x,s) \in [0,1]\times \P \,|\, (\lim_{n\to \infty} \frac{1}{\lambda(x) n} \int_{ \{ g_k^{-1}(x)\cdot 0, g_k^{-1}(x)\cdot \infty \}_G } f_i(z)\, dz )_{i=1\ldots,g} =\alpha \in \R^{2g} \}. $$
For $(x,s)\in E_\alpha$ we have
$$ \lim_{n\to \infty} \frac{1}{\lambda(x) n}  \{ g_k^{-1}(x)\cdot 0, g_k^{-1}(x)\cdot \infty \}_G = h_\alpha \in H_1(X_G,\R), $$
where the homology class $h_\alpha$ is uniquely determined by the property that
$\langle f_i, h_\alpha\rangle =\int_{h_\alpha} f_i(z) dz = \alpha_i$. 

\smallskip

The main result of \cite{KesStra} shows that for a given finite index subgroup $G\subset \GL_2(\Z)$
with $X_G$ of genus $g\geq 1$, there is a strictly convex and differentiable function $\beta_G: \R^{2g}\to \R$
such that, for all $\alpha\in \nabla \beta_G(\R^{2g})\subset \R^{2g}$
\begin{equation}\label{KSresult1}
\dim_H(E_\alpha) =\hat \beta_G(\alpha) ,
\end{equation}
where $\hat \beta_G(\alpha) =\inf_{v\in \R^{2g}} (\beta_G(v) - \langle \alpha, v \rangle)$ is the
Legendre transform, 
and for all $(x,s) \in E_\alpha$
\begin{equation}\label{KSresult2}
 \lim_{n\to \infty} \frac{1}{\lambda(x) n}  \{ g_k^{-1}(x)\cdot g \cdot 0, g_k^{-1}(x)\cdot g\cdot \infty \}_G = h_\alpha(x,s) 
\end{equation}
with $g$ a representative of the class $s\in \P=\GL_2(\Z)/G$.

\smallskip

Let $E=\{ E_\alpha \}$ be the collection of the level sets $E_\alpha$ of the limiting
modular symbol. 
We let $\cA_{\partial,G,\cP,h_\alpha}=
B_{\partial, E}\rtimes {\rm Red}$ be the algebra associated to the collection $E$ of invariant sets, 
as in Corollary~\ref{BEsubalg}.
As an immediate consequence of the results \eqref{KSresult1}, \eqref{KSresult2} 
of \cite{KesStra} we have the following.

\begin{lem}\label{SG2linT}
The choice of a cusp form $\psi\in \cS_{G,2}$ determines a bounded linear operator $\cI_{\psi,\alpha}$ from
$\cA_{\GL_2(\Z),G,\cP}$ to $B_{\partial, E_\alpha}$ with for $(x,s)\in E_\alpha$
\begin{equation}\label{Iop}
\begin{array}{rl}
\displaystyle{ \cI_{\psi,\alpha}(f)(g,\rho,x,s)} =&\displaystyle{ \int_{ \{ \star, x \}_G } f(g,\rho,z,s) \, \psi(z)\, dz} \\[3mm] 
=& \displaystyle{
    \lim_{n\to \infty} \frac{1}{\lambda(x)\, n}\sum_{k=1}^n \int_{ \{ g_k^{-1}(x)\cdot 0, g_k^{-1}(x)\cdot \infty \}_G }  f(g,\rho,z,s) \, \psi(z)\, dz} \\[3mm]
= & \displaystyle{ \int_{h_\alpha(x)} f(g,\rho,z,s) \, \psi(z)\, dz . }    
     \end{array}
\end{equation}
\end{lem}

Here we pair the form $\omega(z)= f(z)\psi(z)dz$ with the limiting modular symbol $h_\alpha(x,s)$
of \eqref{KSresult2}. We use the notation $\omega(z)=\omega_{\rho,s}(z)$
to highlight the dependence on the variables $(\rho,s)$ that come from choosing an element $f$ in the arithmetic
algebra $\cA^{ar}_{\GL_2(\Z),G,\cP}$. 

\smallskip

Note that $\cI_{\psi,\alpha}$ is only a linear operator and {\em not} an algebra homomorphism. 
We obtain a subalgebra of $\cA_{\partial,G,\cP}$ as follows.

\smallskip

\begin{defn}\label{Ialg}
Let $\cA_{\cI,G,\cP}$ denote the subalgebra of $\cA_{\partial,G,\cP}=B_\partial
\rtimes {\rm Red}$ generated by all the images $\cI_{\psi,\alpha}(f)$ for $f\in \cA_{\GL_2(\Z),G,\cP}$,
for $\psi\in \cS_{G,2}$, and for $\alpha \in \nabla \beta_G(\R^{2g})$, and by the $S_k, S_k^*$
with the relations as in Definition~\ref{semicrossdef}. The arithmetic
algebra $\cA^{ar}_{\cI,G,\cP}$ is obtained in the same way as the algebra generated by
the images $\cI_{\psi,\alpha}(f)$ with $f$ in the arithmetic algebra $\cA^{ar}_{\GL_2(\Z),G,\cP}$
of \S \ref{arithmSec},
for all $\psi\in \cS_{G,2}$ and $\alpha \in \nabla \beta_G(\R^{2g})$, and by the $S_k,S_k^*$ as
above.
\end{defn}

\medskip
\subsection{Evaluation of ground states on boundary values}

When we evaluate zero-temperature KMS states on the elements $\cI_{\psi,\alpha}(f)$, for
an element $f\in \cA^{ar}_{\GL_2(\Z),G,\cP}$, we
obtain the pairing of a cusp form with a limiting modular symbol,
\begin{equation}\label{KMSIpsi}
 \varphi_{\infty,\rho,x,s}(\cI_{\psi,\alpha}(f)) =\langle  \omega_{\rho,s}, h_\alpha(x) \rangle, 
\end{equation} 
where $\omega_{\rho,s}(z)=f(1,\rho,z,s) \psi(z) dz$ is a cusp form in $\cS_{G,2}$ for all $(\rho,s)$.  
Since elements $f\in \cA^{ar}_{\GL_2(\Z),G,\cP}$
depend on the variable $\rho\in M_2(\hat\Z)$ through some finite projection $\pi_N(\rho)\in \Z/N\Z$,
we can write $\omega_{\rho,s}(z)$ as a finite collection $\{ \omega_{i,s}(z) \}_{i\in \Z/N\Z}$. 

\smallskip

To illustrate the properties of the values of ground states on arithmetic
boundary elements, we consider here the particular case where $G=\Gamma_0(N)$
and a state $\varphi_{\infty,\rho,x,s}$ with $s\in \P$. 
We also choose $f$ and $\psi$ so that the resulting $\omega_{i,s}$ are 
cusp forms for $\Gamma_0(N)$ that are Hecke eigenforms for all the Hecke
operators $T(m)$.

\smallskip

Recall (see \cite{Serre}) that the Hecke operators $T(m)$ given by 
$$ T_m=\sum_{\gamma\,:\, \det(\gamma)=m} \Gamma_0(N) \gamma \Gamma_0(N) $$
satisfying $T_nT_m=T_mT_n$ for $(m,n)=1$ and $T_{p^n} T_p=T_{p^{n+1}} + p T_{p^{n-1}} R_p$
with $R_\lambda$ the scaling operator that acts on a modular form of weight $2k$ as
multiplication by $\lambda^{-2k}$. The Hecke operators $T_m$ and the scaling operators $R_\lambda$ 
generate a commutative algebra, and the action of $T_m$ on a modular form of weight $2k$ is
given by
$$ T_m\, f(z) = n^{2k-1} \sum_{a\geq 1, ad=n, 0\leq b <d} d^{-2k} \, f(\frac{az+b}{d}). $$

\smallskip

\begin{prop}\label{evalKMSinftyMerel}
Let $G=\Gamma_0(N)$ and let $\omega_{i,s_0}$, with $i=0,\ldots, N-1$ and $s_0=\Gamma_0(N) g_0 \in \P$, 
be Hecke eigencuspforms of weight $2$. For $s\in \P$ with $s=\Gamma_0(N) g_0 \gamma$, with 
$\gamma\in \GL_2(\Z)$, let $\omega_{i,s}=\omega_{i,s_0 \gamma}:=\omega_{i,s_0}\circ \gamma^{-1}$. 
Consider the pairing 
$$ \xi_\omega(s)= \langle  \omega_{\rho,s}, h_\alpha(x) \rangle =\int_{h_\alpha(x)} \omega_{i,s} $$
with the limiting modular symbol $h_\alpha(x)$, as in \eqref{Iop} and \eqref{KMSIpsi}. For $(m,N)=1$
we have the relations
$$ a_{i,m} \,  \xi_\omega(s) = \sum_{M \in \bar A_m} u_m\, \xi_\omega(s\, M) $$
where $a_{i,m}$ are the Hecke eigenvalues and 
$A_m=\{ M \in M_2(\Z)\,:\, \det(M)=m \}$, with $\bar A_m =A_m/\{ \pm 1\}$ and
$\sum_M u_M\, M \in \Z \bar A_m$ is the Manin--Heilbronn lift of the Hecke operator $T_m$.
\end{prop}

\proof The condition that $\omega_{i,s \gamma}=\omega_{i,s}\circ \gamma^{-1}$ implies
that 
$\langle  \omega_{i,s}, h_\alpha(x) \rangle=\langle \omega_{i,s_0}, h_\alpha(x,s) \rangle$.
The following facts are known from \cite{Man-modsymb}, \cite{Merel}.  Let 
$$ A_{m,N}=\{ M=\begin{pmatrix} a & b \\ c & d \end{pmatrix} \in M_2(\Z)\,:\, \det(M)=m, \, \, N|c  \}. $$
Let $R$ be a set of representatives for the classes $\Gamma_0(N)\backslash A_{m,N}$
The Hecke operators act on the modular symbols by 
$T_m \{ \alpha, \beta \}=\sum_{\lambda \in R} \{ \lambda \alpha, \lambda \beta \}$. 
For $(m,N)=1$ there is a bijection between the cosets $\Gamma_0(N)\backslash A_{m,N}$
and $A_m/\SL_2(\Z)$. For $s\in \P$ consider the assignment $\xi_\omega: s\mapsto \xi_\omega(s)= \langle \omega_{i,s}, h_\alpha(x,s)\rangle$,
where $\omega_{i,s}$ is a Hecke eigencuspform. It is shown in  \cite{Man-modsymb}, \cite{Merel}
that there is a lift $\Theta_m$ of the action of the Hecke operators $T_m \circ \xi=\xi\circ \Theta_m$ 
(the Manin--Heilbronn lift), which is given by $\Theta_m =\sum_{\gamma \in A_m /\SL_2(\Z)} \Upsilon_\gamma$,
where $\Upsilon_\gamma$ is a formal chain of level $m$ connecting $\infty$ to $0$ and of class $\gamma$.
This means that $\Upsilon_\gamma=\sum_{k=0}^{n-1} \gamma_k$ in $\Z A_m$, for some $n\in \N$ where
$$ \gamma_k =\begin{pmatrix} u_k & u_{k+1} \\ v_k & v_{k+1} \end{pmatrix} $$
with $u_0/v_0=\infty$ and $u_n/v_n=0$ and where $\gamma_k$ agrees with $\gamma$ 
in $A_m/\SL_2(\Z)$. An argument in \cite{Merel} based on the continued fraction expansion 
and modular symbols shows that it is always possible to construct such formal chains with $\gamma_k =\gamma g_k$ with $g_k\in \SL_2(\Z)$ and
that the resulting $\Theta_m$ is indeed a lift of the Hecke operators. (The length $n$ of the chain
of the Manin--Heilbronn lift is also computed, see \S 3.2 of \cite{Merel}.) Using the notation of
Theorem~4 of \cite{Merel}, we write the Manin--Heilbronn lift as 
$\Theta_m = \sum_{M\in \bar A_m} u_M\, M$ as an element of $\Z A_m$. Each element 
$M\in \bar A_m$ maps $s\mapsto s\, M$ in $\cP$, hence one obtains a map $\Theta_m: \Z\cP \to \Z\cP$. 
In particular, as shown in Theorem~2 of \cite{Merel}, for $s=\Gamma_0(N)g$ in $\P$ one has
$\Theta_m(s)=\sum_{\gamma\in R} \sum_{k=0}^{n-1} \phi(g\gamma \gamma_k)$ where 
$\phi: A_m \to \Gamma_0(N)\backslash \SL_2(\Z)$ is the map that assigns 
$$ A_m \ni \gamma=\begin{pmatrix} a & b \\ c & d \end{pmatrix} \mapsto \Gamma_0(N) 
\begin{pmatrix} w & t \\ u & v \end{pmatrix} \in \Gamma_0(N) \backslash \SL_2(\Z) $$
with $(c :d)=(u :v)$ in $\P^1(\Z/N\Z)\simeq \P =\Gamma_0(N) \backslash \SL_2(\Z)$.
Thus, as in Theorem~2 of \cite{Merel} we then have 
$\Theta_m(s)=\sum_{\gamma\in R} \sum_{k=0}^{n-1} \phi(g\gamma) g_k=\sum_{\gamma\in R} \phi(g\gamma) \gamma^{-1}$, seen here as an element in $\Z\cP$. Thus, in the pairing of limiting modular symbols and
boundary elements we find
$$ T_m \xi_\omega(s) =T_m \int_{h_\alpha(x,s)} \omega_{i,s_0} =\int_{h_\alpha(x,s)} T_m \omega_{i,s_0}
= a_{i,m} \int_{h_\alpha(x,s)} \omega_{i,s_0} ,$$
where $a_{i,m}$ are the Hecke eigenvalues of the eigenform $\omega_{i,s_0}$, with $a_{i,1}=1$.
On the other hand, using the Manin--Heilbronn lift we have
$$ T_m \xi_\omega(s) = \xi_\omega (\Theta_m(s)) =
\sum_{\gamma\in R} \int_{h_\alpha(x,s_\gamma)} \omega_{i,s_0} $$
with $s_\gamma \in \cP$ given by $s_\gamma=\phi(g\gamma) \gamma^{-1}$ as above.
We write the latter expression in the form 
$$ \sum_{M\in \bar A_m} u_M\, \int_{h_\alpha(x,s\, M)} \omega_{i,s_0} $$
for consistency with the notation of Theorem~4 of \cite{Merel}. 
\endproof

\smallskip

\begin{prop}\label{L1period}
Under the same hypothesis as Proposition~\ref{evalKMSinftyMerel}, 
let $L_{\omega_{i,s_0}}(\sigma)=\sum_m a_{i,m} \, m^{-\sigma}$ be the $L$-series associated to the
cusp form $\omega_{i,s_0}=\sum_m a_{i,m} \, q^m$. For $x$ a quadratic irrationality 
the evaluation \eqref{KMSIpsi} of KMS$_\infty$ states satisfies
\begin{equation}\label{L1eval}
 \langle \omega_{i,s}, h(x) \rangle=
 \frac{1}{\lambda(x) n} \sum_{k=1}^n \langle \omega_{i,s_k}, \{0, \infty\} \rangle
= \frac{1}{\lambda(x) n} \sum_{k=1}^n L_{\omega_{i,s_k}}(1),
\end{equation}
where $s_k=\Gamma_0(N) g g_k^{-1}(x)$ for $s=\Gamma_0(N) g$.
\end{prop}

\proof
The special value $L_{\omega_{i,s_0}}(1)$ of the $L$-function gives the pairing with the
modular symbol $\langle \omega_{i,s_0}, \{ 0, \infty\} \rangle$. Similarly, for $s\in \P$
with $s=\Gamma_0(N)g$, the special value gives $$L_{\omega_{i,s}}(1)=\langle \omega_{i,s}, \{ 0, \infty\} \rangle
=\langle \omega_{i,s_0}, \{ g\cdot 0, g\cdot \infty\} \rangle.$$ 
In the case of a quadratic irrationality the limiting modular symbol satisfies
$$ h(x)=\frac{1}{\lambda(x) n} \sum_{k=1}^n \{ g_k^{-1}(x)\cdot 0, g_k^{-1}(x) \cdot \infty \}_G, $$
where $n$ is the length of the period of the continued fraction expansion of $x$ and $\lambda(x)$
is the Lyapunov exponent. For $s_k=\Gamma_0(N) g g_k^{-1}(x)$ with $s=\Gamma_0(N) g$, we have
$\langle \omega_{i,s_k}, \{0, \infty\} \rangle = \langle \omega_{i,s}, \{ g_k^{-1}(x)\cdot 0, g_k^{-1}(x) \cdot \infty \}\rangle$ hence one obtains \eqref{L1eval}.
\endproof

\smallskip

 As shown in Theorem~3.3 of
\cite{Man-modsymb}, the special value $L_{\omega_{i,s_0}}(1)$ satisfies 
$$ (\sum_{d|m} d - a_{i,m}) L_{\omega_{i,s_0}}(1) = \sum_{d|m, \, b \, {\rm mod} \, d} 
\int_{\{ 0, b/d \}_G} \omega_{i,s_0}, $$
since one has
$$ \int_0^{\infty} T_m \omega_{i,s_0} = a_{i,m} \int_0^{\infty} \omega_{i,s_0} =
\sum_{d|m} \sum_{b=0}^{d-1} \int_{\{ b/d,0 \}_G +\{ 0, \infty \}_G} \omega_{i,s_0}. $$

\smallskip

For a normalized Hecke eigencuspform $f=\sum_n a_n q^n$, let 
$L_f(s)=\sum_n a_n n^{-s}$ be the associated $L$-function and
$\Lambda_f(s)=(2\pi)^{-s}\Gamma(s) L_f(s)$ the completed $L$-function,
the Mellin transform $\Lambda_f(s)=\int_0^\infty f(iz) z^{s-1} dz$.

\smallskip

The relation between special values of $L$-functions and periods of Hecke
eigenforms generalizes for higher weights, and it was shown in \cite{Man-Hecke}
that ratios of these special values of the same parity are algebraic (in the field
generated over $\Q$ by the Hecke eigenvalues).
For a normalized Hecke eigencuspform $f=\sum_n a_n q^n$ of weight $k$ 
the coefficients of the period polynomial $r_f(z)$ are expressible in terms of special
values of the $L$-function,
$$ r_f(z)=-i \sum_{j=0}^{k-2} \binom{k-2}{j} (iz)^j \Lambda_f(j+1). $$
Manin's Periods Theorem shows that, for $\K_f$ the field of algebraic numbers 
generated over $\Q$ by the Fourier coefficients, there are $\omega_\pm(f)\in \R$
such that for all $1\leq s \leq k-1$ with $s$ even $\Lambda_f(s)/\omega_+(f)\in \K_f$,
respectively $\Lambda_f(s)/\omega_-(f)\in \K_f$ for $s$ odd.

\smallskip

Shokurov gave a geometric argument based on Kuga varieties and a higher
weight generalization of modular symbols, \cite{Shok}.  It is expected that the
limiting modular symbols of \cite{ManMar}, as well as the quantum statistical mechanics 
of the $\GL_2$-system and its boundary described here, will generalize to the case of
Kuga varieties, with the relations between periods of Hecke eigencuspforms described
in \cite{Man-Hecke} arising in the evaluation of zero-temperature KMS states of these
systems.

\bigskip

\subsection*{Acknowledgment} The first author was partially supported by 
NSF grants DMS-1707882 and DMS-2104330, by NSERC Discovery Grant RGPIN-2018-04937 
and Accelerator Supplement grant RGPAS-2018-522593. 

\subsection*{Conflict of Interest} 
The authors have no conflicts to disclose.

\end{document}